\pdfoutput=1

\documentclass[aps,prd,twocolumn,preprintnumbers,superscriptaddress,nofootinbib,floatfix]{revtex4-1}

\usepackage{bm}

\usepackage[hyperfootnotes=false]{hyperref}
\usepackage{url}

\usepackage{color}
\usepackage[usenames,dvipsnames,svgnames,table]{xcolor}

\usepackage{amsmath}
\usepackage{amssymb}
\usepackage{graphicx}
\usepackage{slashed}
\usepackage{soul}
\usepackage{mathrsfs}

\newcommand{\sell}{\widetilde{\ell}}
\newcommand{\stau}{\widetilde{\tau}}
\newcommand{\smu}{\widetilde{\mu}}
\newcommand{\se}{\widetilde{e}}

\newcommand{\ME}{\slashed{E}}

\usepackage{array}
\newcolumntype{L}[1]{>{\raggedright\let\newline\\\arraybackslash\hspace{0pt}}m{#1}}
\newcolumntype{C}[1]{>{\centering\let\newline\\\arraybackslash\hspace{0pt}}m{#1}}
\newcolumntype{R}[1]{>{\raggedleft\let\newline\\\arraybackslash\hspace{0pt}}m{#1}}

\begin{document}

\title{Hunting for Scalar Lepton Partners at Future Electron Colliders} 

\newcommand{\OKC}{\affiliation{The Oskar Klein Centre for Cosmoparticle Physics, Department of Physics, \\Stockholm University, Alba Nova, 10691 Stockholm, Sweden}}
\newcommand{\SITP}{\affiliation{Stanford Institute for Theoretical Physics, Department of Physics, Stanford University, Stanford, CA 94305, USA}}
\newcommand{\Utah}{\affiliation{Department of Physics and Astronomy, University of Utah, Salt Lake City, UT 84112, USA}}

\author{Sebastian~Baum}
\email{sbaum@stanford.edu}
\SITP
\OKC

\author{Pearl~Sandick}
\email{sandick@physics.utah.edu}
\Utah

\author{Patrick~Stengel}
\email{patrick.stengel@fysik.su.se}
\OKC

\begin{abstract}
New physics close to the electroweak scale is well motivated by a number of theoretical arguments. However, colliders, most notably the Large Hadron Collider (LHC), have failed to deliver evidence for physics beyond the Standard Model. One possibility for how new electroweak-scale particles could have evaded detection so far is if they carry only electroweak charge, i.e. are color neutral. Future $e^+e^-$ colliders are prime tools to study such new physics. Here, we investigate the sensitivity of $e^+e^-$ colliders to scalar partners of the charged leptons, known as sleptons in supersymmetric extensions of the Standard Model. In order to allow such scalar lepton partners to decay, we consider models with an additional neutral fermion, which in supersymmetric models corresponds to a neutralino. We demonstrate that future $e^+e^-$ colliders would be able to probe most of the kinematically accessible parameter space, i.e. where the mass of the scalar lepton partner is less than half of the collider's center-of-mass energy, with only a few days of data. Besides constraining more general models, this would allow to probe some well motivated dark matter scenarios in the Minimal Supersymmetric Standard Model, in particular the {\it incredible bulk} and {\it stau co-annihilation} scenarios.
\end{abstract}

\maketitle

\section{Introduction} 
The hierarchy problem and the WIMP (Weakly Interacting Massive Particle) paradigm both point to new physics close to the electroweak scale. However, after nearly a decade of operations, the Large Hadron Collider (LHC) has not delivered evidence for physics Beyond the Standard Model (BSM). While it is possible that the scale of new physics is beyond the reach of the LHC, it is also possible that new electroweak-scale particles are lurking in a blind spot of the LHC. 

Prime examples of models that could have thus far evaded detection include those where the new states are charged only under electromagnetism and not under the strong force (Quantum Chromodynamics, QCD). While such new states may still have sizable production cross sections at hadron colliders, the signal can easily be drowned out by SM QCD processes. The background conditions at hadron colliders are particularly challenging when the BSM scenario yields only relatively soft visible decay products, as is the case for compressed BSM spectra where several new particles are nearly degenerate in mass. Lepton colliders could probe such BSM scenarios much more easily for two reasons: First, leptons are color neutral and thus backgrounds at lepton colliders are produced via electroweak processes. Second, leptons are fundamental particles, rendering collisions at lepton colliders much cleaner than events at hadron colliders where composite protons interact.

A number of future $e^+e^-$ colliders have been proposed, such as the International Linear Collider (ILC)~\cite{Behnke:2013xla,Aihara:2019gcq}, the Compact Linear Collider (CLIC)~\cite{Charles:2018vfv}, the Circular Electron Positron Collider (CEPC)~\cite{CEPCStudyGroup:2018rmc,CEPCStudyGroup:2018ghi}, and the $e^+ e^-$ option of the Future Circular Collider (FCC-ee)~\cite{Abada:2019zxq}. ILC and CLIC are linear colliders, while CEPC and FCC-ee are circular colliders. Center-of-mass energies as low as $\sqrt{s} = 250\,$GeV will suffice for making progress in electroweak and Higgs precision physics, notably via production of Higgs bosons in association with $Z$ bosons. New physics hiding in LHC blind spots could be evident at low center-of-mass energies, but, generally, larger $\sqrt{s}$ is more advantageous for new physics searches. The ILC has been proposed with a $\sqrt{s} = 500\,$GeV option~\cite{Behnke:2013xla,Aihara:2019gcq}, and CLIC may reach energies as large as $\sqrt{s} = 3\,$TeV~\cite{Charles:2018vfv}. All proposed collider projects are envisaged to collect integrated luminosities of a few ab$^{-1}$. 

In this work, we examine the prospects of future lepton colliders to probe scenarios where the SM is extended by scalar partners of the charged leptons. In supersymmetric (SUSY) models, such particles are referred to as sleptons. We will use SUSY-inspired language to refer to such particles in the remainder of this paper, but our results apply to a wide class of new spin zero particles charged only under electromagnetism, as we will see below. More concretely, we consider the pair production of such sleptons and their subsequent decay to the associated charged SM lepton and a new neutral fermion which leaves the detector without depositing energy. While the sensitivity of lepton colliders to new fermionic states charged only under the electroweak force, referred to as electroweakinos in SUSY language, has been studied extensively, see e.g. Refs.~\cite{Fujii:2015jha,CEPCStudyGroup:2018ghi,dEnterria:2016sca,deBlas:2018mhx,Baer:2019gvu,Habermehl:2020njb} and references therein, sleptons have received considerably less attention (note important work in~\cite{Farrar:1980uy,Tsukamoto:1993gt,Nojiri:1994it,Feng:2001ce,Freitas:2002gh,Boos:2003vf,Freitas:2003yp,Martyn:2004jc,Battaglia:2005zf,Buckley:2007th,Ellis:2008ca,Bechtle:2009em,Berggren:2013vna,Endo:2013lva}).

We consider BSM models extending the SM by a slepton $\sell$, a spin-zero state with electromagnetic charge and lepton number one, but neutral under QCD. If we assume lepton number and flavor to be conserved, the slepton has no decay mode that includes only SM particles. Colliders constrain the mass of new stable charged scalars to be $m_{\sell} \gtrsim 450\,$GeV~\cite{Khachatryan:2016sfv,Aaboud:2019trc}. Charged stable particles are also severely constrained by cosmology, see e.g. Refs.~\cite{Pospelov:2006sc,Pospelov:2008ta, Burdin:2014xma}. In order to allow for the slepton to decay in a lepton number and flavor conserving manor, we introduce a neutral fermion $\chi$ which does not carry lepton number. For $m_{{\chi}} < m_{\sell} + m_\ell$, this opens the decay channel $\sell \to \ell + \chi$, where $\ell$ is the charged SM lepton of the same flavor as $\sell$. If $\sell$ carries muon number, we will refer to it as a smuon $\smu$, if it carries tau number as a stau $\stau$, and if it carries electron number as a selectron $\se$. 

SUSY models such as the Minimal Supersymmetric Standard Model (MSSM)~\cite{Nilles:1983ge,Martin:1997ns,Chung:2003fi} provide concrete realizations of BSM models with sleptons $\sell$ and neutral fermions ${\chi}$. More specifically, in SUSY models the supersymmetric partners of the leptons (i.e. the sleptons) play the role of $\sell$, and the partners of the neutral Higgs bosons and the electroweak gauge bosons give rise to fermionic neutral states (i.e. the neutralinos) playing the role of $\chi$ (see e.g. Ref.~\cite{Fayet:1977yc}). Instead of lepton number conservation invoked above, in SUSY models $R$-parity~\cite{Farrar:1978xj} ensures the stability of the Lightest Supersymmetric Particle (LSP). For our purposes, $R$-parity can be understood as a $\mathbb{Z}_2$ symmetry under which all SM states are even, and their superpartners (i.e. the new states) are odd. If one of the neutralinos is the LSP, $R$-parity thus renders $\chi$ a dark matter candidate. Models where one of the smuons or staus is the next-to-LSP (NLSP) are of particular interest for dark matter model building in the MSSM because they offer a way to achieve the correct dark matter relic density, see e.g. Refs.~\cite{Ellis:1998kh,Ellis:1999mm,Buckley:2013sca,Pierce:2013rda,Fukushima:2014yia,Baker:2018uox,Duan:2018cgb}

Sleptons can be pair produced at colliders, where they would then decay into a pair of opposite-sign charged SM leptons of the same flavor as the sleptons and a pair of $\chi$'s. Since the $\chi$ is, by construction, neutral and stable, it would leave the detector without depositing energy, manifesting as {\it missing energy}. For smuons and selectrons, the collider signature will thus be a pair of opposite-sign same-flavor leptons plus missing energy. For staus, a number of final states arise depending on the decay mode of the taus. Leptonic tau decays give rise to particularly clean final states, and, as we will see, the most interesting final state will be a pair of opposite-sign leptons of different flavor accompanied by missing energy.

Prospects for slepton searches at hadron colliders such as the LHC have been discussed extensively in the context of the MSSM (see e.g. Refs.~\cite{Aad:2014vma,Dutta:2014jda,Han:2014aea,Aaboud:2017leg,Aaboud:2018jiw,Sirunyan:2018nwe,Dutta:2017nqv,Fuks:2019iaj}). The ability to probe different parts of the parameter space depends crucially on the spectrum of SUSY particles, in particular the slepton mass $m_{\sell}$ and the mass gap between the slepton and the neutralino, $m_{\sell} - m_{\chi}$. It is relatively straightforward to probe large mass gaps ($m_{\sell }- m_{\chi} \gtrsim 60\,$GeV) at the LHC. For such mass gaps, the LHC will have discovery reach for slepton masses as large as $m_{\sell} \lesssim 500\,$GeV with $L = \mathcal{O}(500)\,{\rm fb}^{-1}$ of data~\cite{Fuks:2019iaj}. However, the kinematics make it difficult to differentiate the signal from the SM background coming from leptonic decays of $W^+ W^-$ pairs for $m_{\sell} - m_{\chi} \sim m_W/2$. In order to improve the sensitivity of the LHC to more compressed scenarios (i.e. smaller mass gaps $m_{\sell} - m_{\chi}$), one can search for slepton pair production accompanied by an additional hard (hadronic) jet from initial state radiation (ISR). Such an ISR jet provides a transverse boost to the $\sell^+\sell^-$ system. Using kinematic cuts taking advantage of the boosted topology, one can regain some sensitivity to the compressed region. Reference~\cite{Han:2014aea} focused on the regime $5\,{\rm GeV} \lesssim m_{\sell} - m_{\chi} \lesssim 20\,$GeV, and showed that the LHC could probe this compressed region for slepton masses up to $m_{\sell} \lesssim 150\,$GeV. Reference~\cite{Dutta:2014jda} obtained similar reach for the compressed regime using hadronic jets from vector boson fusion production of the sleptons instead of an ISR jet. Reference~\cite{Dutta:2017nqv} investigated the ability of the LHC to probe intermediated mass gaps, $10\,{\rm GeV} \lesssim m_{\sell} - m_\chi \lesssim 60\,$GeV, using a hard ISR jet. Ref.~\cite{Dutta:2017nqv} found that with the data expected to be collected in upcoming LHC runs, slepton masses up to $m_{\sell} \lesssim 200\,$GeV could be probed for mass gaps $10\,{\rm GeV} \lesssim m_{\sell} - m_\chi \sim 20\,$GeV, and that mass gaps larger than that are more challenging to probe because of the kinematic similarity to the SM $W^+W^-$ background as $m_{\sell} - m_\chi \to m_W/2$. Note that Refs.~\cite{Dutta:2014jda,Han:2014aea,Dutta:2017nqv} focused on selectrons and smuons; the LHC is even less sensitive to staus, because ($\stau^\pm \to \tau^\pm \chi$) decays give rise to softer charged leptons in the final states than ($\se^\pm/\smu^\pm \to e^\pm/\mu^\pm + \chi$) decays.

Compared to hadron colliders, slepton searches at $e^+e^-$ colliders are considerably more straightforward. Because of the absence of the underlying event and pile-up, which clutter the final states at hadron colliders, and because at an $e^+e^-$ collider the detector frame coincides with the center-of-mass frame of the $e^+e^-$ collisions, reconstructing soft charged leptons as well as missing energy is much simpler. Furthermore, future $e^+e^-$ colliders are envisaged to operate without triggers, hence, even final states containing only very soft leptons and relatively small missing energy can be used for physics searches.

Searches for selectron and smuon pair production at $e^+e^-$ colliders were first proposed in Ref.~\cite{Farrar:1980uy}, while searches for stau pair production were discussed first in Refs.~\cite{Tsukamoto:1993gt, Nojiri:1994it}. Most previous studies investigating slepton pair production at future $e^+e^-$ colliders focused on the reconstruction of the mass spectra and values of the couplings for particular MSSM benchmark points, see e.g. Refs.~\cite{Tsukamoto:1993gt, Nojiri:1996fp, Feng:2001ce, Freitas:2002gh, Boos:2003vf, Freitas:2003yp, Martyn:2004jc, Battaglia:2005zf, Buckley:2007th, Bechtle:2009em, Berggren:2013vna,Endo:2013lva}. The discovery potential of lepton colliders within the Constrained MSSM (CMSSM) and similar scenarios has been discussed in Ref.~\cite{Ellis:2008ca}. A study of the coverage of SUSY scenarios with a smuon or stau NLSP and neutralino LSP at future lepton colliders for a wide range of LSP and NLSP masses has been presented in Ref.~\cite{Berggren:2013vna} (the results of this study have also been re-published in Refs.~\cite{Baer:2013vqa, Berggren:2015qua}), with the aim to determine the regions of the LSP-NLSP mass plane within reach of a $\sqrt{s} = 500\,$GeV $e^+e^-$-collider. 

In this work we present a more general scan of the parameter space of a simplified model with light sleptons. We consider a spectrum where only one smuon or stau is close to the mass of the neutral fermion $\chi$ and assume the slepton decays exclusively to $\chi$ and the SM lepton of the same flavor as the slepton. Such an approach allows us to explore the sensitivity of future $e^+e^-$ colliders to general BSM models containing scalar lepton partners and new neutral fermions in a broad setting. While our results can directly be applied to SUSY models, they can also be interpreted in more general models, without the particular relations between model parameters introduced by SUSY. The kinematic cuts we use for event selection are qualitatively similar to those used in previous studies, however, in our case the cuts are optimized for a wide region of masses of the slepton and $\chi$ instead of a particular benchmark point. Furthermore, previous studies discussing searches for stau pair production at $e^+e^-$ colliders made use of semi-hadronic or fully hadronic decays of the taus produced in the stau decays, see e.g. Refs.~\cite{Tsukamoto:1993gt, Nojiri:1994it, Boos:2003vf, Martyn:2004jc, Bechtle:2009em, Berggren:2013vna}. Here, we instead propose to use fully leptonic tau decays, in particular, the final state containing a pair of different-flavor charged leptons accompanied by missing energy, ($e^\pm \mu^\mp + \ME$). While the corresponding tau branching ratios are smaller than the hadronic branching ratios of taus, this final state offers both excellent $\ME$ reconstruction, because the charged leptons are comparatively easy to reconstruct, and comparatively small SM backgrounds.  

We consider two concrete scenarios for an $e^+e^-$ collider: $\sqrt{s} = 250\,$GeV and $\sqrt{s} = 500\,$GeV. We compute the smallest cross section which could be probed as a function of the masses of the slepton and the new neutral fermion $\chi$, assuming luminosities of $L=0.5\,{\rm ab}^{-1}$ for the $\sqrt{s} = 250\,$GeV scenario and $L = 1\,{\rm ab}^{-1}$ for $\sqrt{s} = 500\,$GeV.\footnote{The luminosities referred to here are the integrated luminosities recorded in a specific polarization configuration. Since we are interested in the effect of longitudinal beam polarization on the sensitivity, we choose these benchmark luminosities to be somewhat lower than the total luminosities envisaged for the different lepton collider projects discussed above.} As we will see, for most combinations of masses, the cross sections which could be probed with these luminosities are orders of magnitude smaller than what we expect in our model. Thus, we also compute how much luminosity would be required to probe a given combination of masses of the slepton and the new neutral fermion. We furthermore explore the effects of longitudinal beam polarization on the sensitivity. Reaching sizable polarization fractions is expected to be rather straightforward at linear colliders, while the feasibility of longitudinal beam polarization at circular colliders is a matter of ongoing debate. As we will demonstrate, achieving significant polarization fractions would make it considerably easier to probe sleptons at a future lepton collider. While we present results only for center-of-mass energies $\sqrt{s} \leq 500\,$GeV, we can readily extrapolate from our results that an $e^+e^-$ collider with larger center-of-mass energy could probe sleptons almost all the way to the pair production threshold $m_{\sell} \lesssim \sqrt{s}/2$ for any mass gap $m_{\sell} - m_\chi$ with a few fb$^{-1}$ of data.

This analysis is directly applicable to a number of dark matter scenarios - simplified, stand-alone, or within the MSSM - that can be constrained simply by the mass of the lightest slepton. In a scenario we refer to as the {\it incredible bulk}, the pair-annihilation of a mostly bino-like neutralino LSP (singlet fermion dark matter candidate) proceeds through $t$-channel exchange of either a smuon~\cite{Fukushima:2014yia} or a stau~\cite{Pierce:2013rda} and can yield the correct relic density for sizable left-right mixing of the sleptons.  As discussed in Ref.~\cite{Fukushima:2014yia}, the dark matter annihilation cross-section in the incredible bulk is maximized when the neutralino mass is similar to that of the lightest slepton. Since the relevant couplings of the bino-like neutralino to the lightest slepton are given by the SM hypercharge of the corresponding lepton, it can be shown that the correct relic density cannot be produced via this mechanism for $m_{\sell} \gtrsim 150\,$GeV.
 
In the more well-studied case of {\it stau co-annihilation}, the lightest stau is nearly degenerate in mass with the neutralino LSP and, thus, annihilation processes involving one or two staus in the initial state must also be accounted for in the calculation of the relic density~\cite{Ellis:1998kh,Ellis:1999mm,Buckley:2013sca,Baker:2018uox,Duan:2018cgb}. While in constrained frameworks such as the CMSSM the observed relic density can only be obtained for stau masses below $\sim 400\,$GeV in the co-annilihilation regime (for example, see Ref.~\cite{Citron:2012fg}), the feasible parameter region is larger when considering more general version of the MSSM. A recent analysis in Ref.~\cite{Duan:2018cgb} aimed to identify the largest stau masses permitted in MSSM scenarios with significant left-right mixing. Depending on how constraints from charge-breaking vacua are implemented, they found that the lightest stau must satisfy $m_{\sell} \lesssim 0.9 - 1.4 \,$TeV to obtain the correct relic density. An $e^+e^-$ collider with higher center-of-mass energy, e.g. CLIC, would be required to fully investigate typical co-annihilation scenarios, but, as we will demonstrate, an $e^+e^-$ collider with $\sqrt{s} = 500\,$GeV could probe most of the feasible parameter space of the incredible bulk in the MSSM.

The remainder of this paper is organized as follows: In Sec.~\ref{sec:model} we describe the BSM model studied here in more detail. In Sec.~\ref{sec:BP_sim} we describe the numerical collider simulations we use to compute the sensitivity of future $e^+e^-$ colliders to sleptons, and, in particular, the set of kinematic cuts we use to reduce the SM background. In Sec.~\ref{sec:sens} we show results for the sensitivity of future $e^+e^-$ colliders with $\sqrt{s} = 250\,$GeV and $\sqrt{s} = 500\,$GeV in the $(m_{\sell},m_\chi)$ plane. We present our conclusions in Sec.~\ref{sec:diss}. Additional details about the kinematic cuts we use for the event selection can be found in the Appendices.

\section{Model Setup} \label{sec:model}
We extend the SM by spin zero partners of the charged leptons with electromagnetic charge $Q = -e$ and to which we assign lepton number one. We denote these spin zero partners according to their flavor, $\sell = \se, \smu, \stau$, and for simplicity we refer to $\se$ as selectrons, $\smu$ as smuons, and $\stau$ as staus throughout the text. We assume the conservation of lepton and baryon number (as well as flavor), which ensures the stability of the lightest slepton.\footnote{Stability of the lightest slepton could also be ensured via matter parity, i.e. assuming $B-L$ to be conserved, instead of conserving $L$ and $B$ individually.} We assume the sleptons to be the $Q=-e$ component of an $SU(2)_L$ multiplet. We limit ourselves to considering sleptons stemming from $SU(2)_L$ singlets and $SU(2)_L$ doublets in this paper, although our results straightforwardly extend to sleptons stemming from larger representations. In analogy to the SM conventions, we refer to sleptons stemming from $SU(2)_L$ singlets as right-handed sleptons $\sell_R$, and states stemming from $SU(2)_L$ doublets as left-handed sleptons $\sell_L$. We assume flavor conservation, forbidding mixing between sleptons of different flavor, but allow for intergenerational mixing, i.e. left-right mixing. We write the mass eigenstates $\sell_i = \sell_{1,2}$ for each flavor as
\begin{equation} \label{eq:slepmix}
   \begin{pmatrix} \sell_1 \\ \sell_2 \end{pmatrix} = \begin{pmatrix} \cos\alpha & \sin\alpha \\ -\sin\alpha & \cos\alpha \end{pmatrix} \begin{pmatrix} \sell_L \\ \sell_R \end{pmatrix} \;,
\end{equation}
where $\alpha$ is the left-right mixing angle and $\sell_1$ is the lighter of the two states ($m_{\sell_1} < m_{\sell_2}$).

In order to give the sleptons decay channels, we add a new fermion $\chi$ with $Q=0$ to the model. If this new fermion does not carry lepton number, it opens a decay channel ($\sell^\pm \to \chi \ell^\pm$) if $m_\chi < m_{\sell} - m_\ell$, where $\ell^\pm$ is the charged SM fermion of the same flavor as $\sell$. The fermion $\chi $, which we refer to as a neutralino, can again stem from a non-trivial representation of $SU(2)_L$, although this does not impact the phenomenology considered here. The relevant terms in the Lagrangian are
\begin{equation}
   \mathcal{L} \supset \left| D_\mu \sell \right|^2 - m_{\sell_i}^2 | \sell_i |^2 + \bar{\chi} \left( i \slashed{D} - m_\chi \right) \chi - \left( \kappa \sell_i^\dagger \bar{\chi} \ell_i + {\rm h.c.} \right) \;,
\end{equation}
where $\kappa$ is a free dimensionless coupling. Note that gauge invariance dictates all interactions of the sleptons and the neutralinos with the SM gauge bosons. We have omitted additional operators which couple pairs of sleptons or neutralinos to the Higgs doublet; they do not lead to any relevant new interactions. 

The assumption of lepton and baryon number conservation renders the lightest state of the sleptons (of a given flavor) and neutralinos stable. An alternative route to achieving such stability and ensuring that the sole decay channels of the sleptons and neutralinos are ($\sell^\pm \to \chi \ell^\pm$) or ($\chi \to \sell^\pm \ell^\mp$) would have been to impose a $\mathbb{Z}_2$ symmetry under which the sleptons and neutralinos are odd while all SM fields are even. Before continuing, let us stress that while we use SUSY-inspired language to refer to the new states in this model, SUSY would not impact our results. This is because the only effect of SUSY on this model would be to relate the coupling $\kappa$ to other couplings in the SUSY model, assuming that all other BSM states are heavier than the sleptons and the neutralino.\footnote{Note that for typical choices of parameters in the MSSM, the charged component of a left-handed slepton doublet is heavier than the {\it sneutrino}, the neutral component of that doublet. Thus, the situation where a neutralino is the lightest and a pure left-handed charged slepton the next-to-lightest supersymmetric particle is difficult to achieve in the MSSM. However, the mass splitting between the sneutrino and the slepton is smaller than the $W$-boson mass; thus, $(\sell^\pm \to \ell^\pm + \chi)$ will still be the dominant decay mode since $\sell^\pm \to \widetilde{\nu} + (W^* \to X)$ decays are suppressed by the off-shell nature of the intermediary $W^\pm$ boson. Here, we denote the sneutrino as $\widetilde{\nu}$, and $X$ stands for the possible SM states arising from the decay of the off-shell $W$-boson (e.g. two quarks or $W^\pm \to \ell^\pm v_\ell$).} Since this coupling gives rise to the sole decay channel of the sleptons/neutralinos, the numerical value of $\kappa$ has no effect on the phenomenology. In SUSY models, $R$-parity would have played the role of the $\mathbb{Z}_2$ symmetry mentioned above, ensuring the stability of the lightest new state instead of lepton number conservation. 

At $e^+e^-$ colliders, sleptons may be readily pair produced via $s$-channel exchange of photons and $Z$ bosons. Selectrons may also be pair produced via $t$-channel exchange of a neutralino. If $m_{\sell} < m_\chi$, the slepton will be stable and thus the collider signature will be the pair production of new stable charged particles. Searches carried out at the LHC rule out such new stable charged particles for $m_{\sell} \lesssim 450\,$GeV, see e.g. Refs.~\cite{Khachatryan:2016sfv,Aaboud:2019trc}. When considered in combination with constraints from the formation of exotic atoms containing the sleptons, which can subsequently alter the light element abundances produced during big bang nucleosynthesis (see e.g. Refs.~\cite{Pospelov:2006sc,Pospelov:2008ta}), such charged particles indeed cannot be completely stable. We thus focus on the $m_\chi < m_{\sell}$ case in the remainder of this work, where the neutralino is a neutral, stable particle and, thus, constitutes a good dark matter candidate. Indeed, a number of SUSY-motivated dark matter models feature a SUSY neutralino as the lightest new state and a slepton as the next-to-lightest, see e.g. Refs.~\cite{Ellis:1998kh,Ellis:1999mm,Buckley:2013sca,Pierce:2013rda,Fukushima:2014yia}. 

\begin{figure}
   \includegraphics[width=\linewidth]{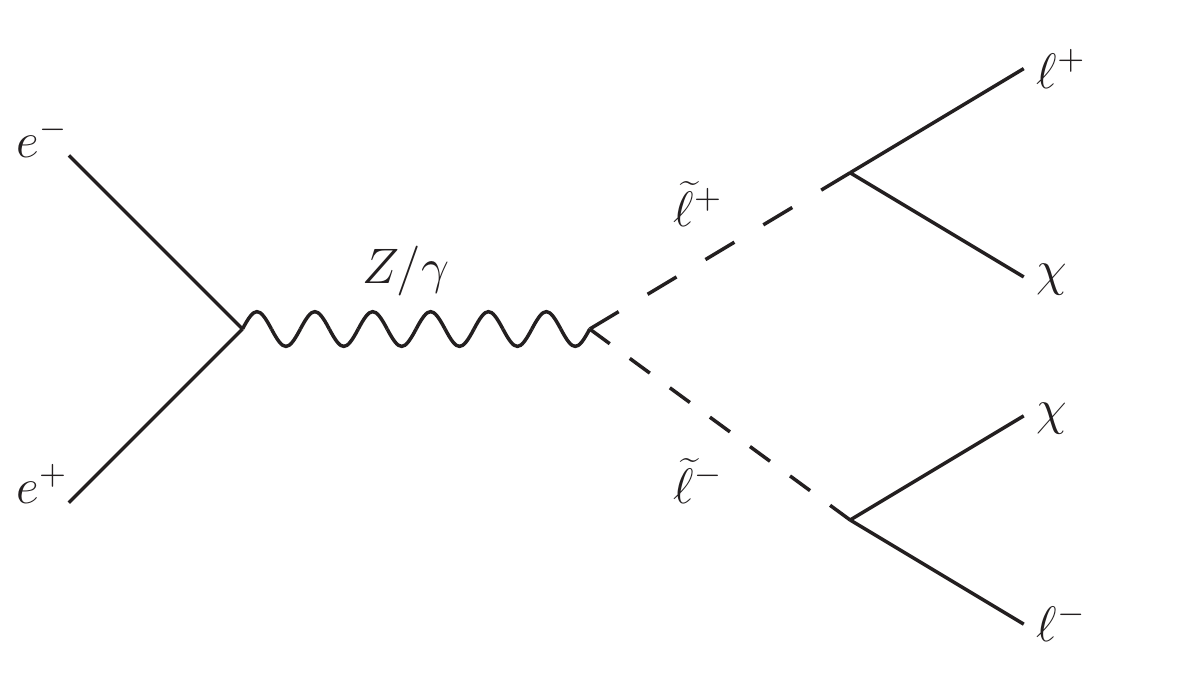}
   \caption{Illustration of slepton pair production via $Z$-boson and photon exchange at a lepton collider and subsequent decays of the sleptons $\sell$ into a charged SM lepton $\ell$ of the same flavor as the slepton and a neutralino $\chi$. Note that for selectrons ($\sell=\se$) there are additional ($e^+ e^- \to \se^+ \se^-$) production diagrams where a neutralino $\chi$ is exchanged in the $t$-channel. We only consider the production of smuons and staus in this work. As discussed in the text, although we use SUSY-inspired language to refer to the states $\sell$ and $\chi$, our analysis is not confined to SUSY models.}
   \label{fig:diagram}
\end{figure}

If $m_\chi < m_{\sell}$, each of the sleptons will decay to a charged SM lepton of the same flavor and a neutralino, as shown in Fig.~\ref{fig:diagram}. Since the neutralinos are neutral and stable, they will leave the detector without depositing energy, manifesting as missing energy $\ME$. In summary, the final states observable at a lepton collider are
\begin{itemize}
   \item for selectrons ($\sell = \se$), a pair of opposite-sign electrons accompanied by missing energy, ($e^+ e^- + \ME$);
   \item for smuons ($\sell = \smu$), a pair of opposite-sign muons accompanied by missing energy, ($\mu^+\mu^- + \ME$);
   \item for staus ($\sell = \stau$), the final state depends on the decay mode of the $\tau^\pm$ produced in the ($\stau^\pm \to \chi \tau^\pm$) decays. The cleanest final states arise from leptonic decays of the taus, resulting in ($e^+ e^- + \ME$), ($e^\pm \mu^\mp + \ME$), and ($\mu^+ \mu^- + \ME$) final states.
\end{itemize}
While interesting, selectron pair production is most difficult to probe because the ($e^+ e^- + \ME$) final state suffers from relatively large backgrounds at $e^+e^-$ colliders. Note that unlike for smuons and staus, the selectron pair production cross section also depends on the value of the coupling $\kappa$. Furthermore, from the standpoint of dark matter model building, achieving sufficiently large annihilation cross sections for canonical freeze-out production requires large left-right mixing of the selectrons, which is generally in conflict with measurements of the electric and magnetic dipole moments of the electron, see e.g. Ref.~\cite{Fukushima:2014yia}. Thus, we focus on smuon and stau production in this paper.

As we will see below, the most challenging background for $\mu^+\mu^- + \ME$ final states comes from $W^+W^-$ pair production (as well as $ZZ$ pair production). At the center-of-mass energies proposed for future $e^+e^-$ colliders, the photon components of the electron/positron beams become non-negligible. This gives rise to sizable ($\gamma\gamma \to \mu^+ \mu^-$) backgrounds which have non-vanishing $\ME$ since the center-of-mass system of $\gamma\gamma$ collisions can feature sizable boosts along the beam axis. For stau searches, the most sensitive final state is ($e^\pm \mu^\mp + \ME$). This channel has much smaller background rates than the final states with a same-flavor lepton pair. For example, compared to the ($\mu^+\mu^- + \ME$) final state, backgrounds from $ZZ$ pair production and ($\gamma\gamma \to \mu^+ \mu^-$) are absent.

Future lepton colliders, in particular linear colliders such as the ILC or CLIC, are envisaged to operate with the option of sizable longitudinal beam polarization. This is advantageous for slepton searches, as the cross sections of both the most relevant background processes and the signal production cross sections are non-trivial functions of the beam polarization. Thus, it is useful to write the production cross section at lepton colliders in terms of the polarized production cross sections 
\begin{equation} \begin{split} \label{eq:xsecpol}
   \sigma &= f_L(e^-) f_L(e^+) \sigma_{LL} + f_L(e^-) f_R(e^+) \sigma_{LR} \\
   &\quad + f_R(e^-) f_L(e^+) \sigma_{RL} + f_R(e^-) f_R(e^+) \sigma_{RR} \;,
\end{split} \end{equation} 
where $\sigma_{ij}$ with $\{i,j\} = \{L,R\}$ is the cross section from $e_i^- e_j^+$ states, and $f_i$ is the (longitudinal) beam polarization fraction
\begin{equation}
   f_L = \frac{n_L}{n_L+n_R}\;, \qquad f_R = \frac{n_R}{n_L+n_R}\;,
\end{equation}
with the number of left-(right-)polarized electrons in the beam $n_L$ ($n_R$). The beam polarization is commonly parameterized by
\begin{equation} \label{eq:PRdef}
   P_R = f_R - f_L \:,
\end{equation}
and we use this convention in the rest of this work. In particular, we denote the polarization as $P = \left[ P_R(e^-), P_R(e^+) \right]$, where $P_R(e^\pm)$ denotes the polarization of the electron/positron beam. In this notation, $P_R = 0.8$, for example, corresponds to beam-polarization fractions $f_R = 90\,\%$ and $f_L = 10\,\%$, while $P_R = -0.8$ corresponds to $f_R = 10\,\%$, $f_L = 90\,\%$. $P_R = 0$ corresponds to an unpolarized beam, $f_L = f_R = 50\,\%$.

For slepton searches, $W^+ W^-$ pair production is one of the most problematic background sources. Because $W$-bosons directly couple to left-handed leptons, but not to right-handed leptons, the production cross section of $W^+ W^-$ pairs can be efficiently suppressed at lepton colliders by choosing mostly right-handed polarizations for the electron and positron beams. The production cross section of left-handed sleptons also falls with increasing right-handed beam polarization, while the production cross section of pairs of right-handed sleptons grows with the right-handedness of the beams.

The production cross section for pairs of smuons or staus can be written as\footnote{Here we are neglecting the photon component of the electron/positron, which can be relevant in highly relativistic $e^+ e^-$ collisions. While properly incorporating the electron's photon component would have negligible effects on the $e^+ e^-$-induced processes discussed in this section, $\gamma\gamma$ induced backgrounds are relevant in our analysis.}

\begin{equation} \begin{split} \label{eq:xsecii}
   \frac{d \sigma_{LL/RR}(e^+e^- \to \sell_i \sell_i)}{d\cos\theta} &= \frac{\pi \alpha_{\rm EM}^2}{2s} \left[ 1 + g_{\sell}^{ii} \; g_{LL/RR} \frac{s}{s-m_Z^2} \right]^2 \\
   & \quad \times \left( 1 - \frac{4 m_{\sell}^2}{s} \right)^{3/2} \sin^2\theta \;,
\end{split} \end{equation}
where $s$ is the squared center-of-mass energy, $\alpha_{\rm EM}$ is the fine structure constant, $m_Z$ is the mass of the $Z$-boson, and $\theta$ is the scattering angle in the center-of-mass frame. The effective coupling of pairs of sleptons to $Z$-bosons is
\begin{align}
   g_{\sell}^{ii} = \begin{cases} \frac{\cos^2\alpha}{\sin(2\theta_W)} - \tan\theta_W &\;{\rm for}\;\sell_i = \sell_1\;, \\ \frac{\sin^2\alpha}{\sin(2\theta_W)} - \tan\theta_W &\;{\rm for}\;\sell_i = \sell_2\;, \end{cases}
\end{align}
where $\sin\theta_W$ is the weak mixing angle and $\alpha$ is the left-right slepton mixing angle, as in Eq.~\eqref{eq:slepmix}. Likewise, the effective coupling of left-(right-)handed electrons to $Z$ bosons is parameterized by
\begin{align}
   g_{LL} &= \frac{1}{2}\left( \cot\theta_W - \tan\theta_W \right) \;, \\
   g_{RR} &= -\tan\theta_W \;.
\end{align}
Smuon and stau pair production proceeds via $s$-channel exchange of a photon or a $Z$-boson. Since neither the photon nor the $Z$-boson couple to combinations of one left-handed and one right-handed electron, $\sigma_{RL} = \sigma_{LR} = 0$.

For non-zero left-right mixing $\sin\alpha \neq 0$, ($e^+ e^- \to \sell_1 \sell_2$) production is possible. Photons do not couple to $\sell_1 \sell_2$, so production proceeds only via $s$-channel exchange of $Z$-bosons. The differential cross section can be written as
\begin{equation} \begin{split} \label{eq:xsec12}
   &\frac{d\sigma_{LL/RR}(e^+e^- \to \sell_1 \sell_2)}{d\cos\theta} \\
      &~ = \frac{\pi \alpha_{\rm EM}^2}{2s} \frac{\sin^2 (2\alpha) g_{LL/RR}^2}{4 \sin^2 (2\theta_W)} \left( \frac{s}{s-m_Z^2} \right)^2 \\
      &\quad \times \left\{ \left[ 1 - \frac{\left(m_1 + m_2 \right)^2}{s} \right] \left[ 1 - \frac{\left(m_1 - m_2 \right)^2}{s} \right] \right\}^{3/2} \sin^2\theta \;,
\end{split} \end{equation}
where $m_1$ and $m_2$ are the masses of $\sell_1$ and $\sell_2$, respectively. Note that these tree-level results are subject to radiative corrections. For the case of the MSSM, Ref.~\cite{Freitas:2003yp} found that radiative corrections typically enhance the smuon and stau pair production cross section by a few percent, except close to the kinematic threshold, where the corrections can be as large as $\sim 10\,\%$. Therefore, our results are relatively robust with respect to such radiative corrections.

The different processes for slepton production at lepton colliders, in particular both the amplitudes mediated by the massive $Z$-boson and the massless photons, all yield the same dependence on the scattering angle
\begin{equation}
   \frac{d\sigma}{d\cos\theta} \propto \sin^2 \theta \;,
\end{equation}
see Eqs.~\eqref{eq:xsecii} and~\eqref{eq:xsec12}. The sole effect of changing the polarization of the electron/positron beams and/or the left-right mixing angle $\alpha$ on the signal spectra is thus to change the normalization, not the shapes of the kinematic distributions.

Finally, the production cross sections of the observable final states can be obtained by taking into account the relevant branching ratios as long as the narrow width approximation holds. For smuons, the ($\mu^+\mu^- + \ME$) final state production cross section is
\begin{equation} \begin{split}
   &\sigma(e^+ e^- \to \smu^+ \smu^- \to \mu^+ \mu^- + \ME) \\
   &= \sigma(e^+ e^- \to \smu^+ \smu^-) \times \left[{\rm BR}(\smu^\pm \to \chi \mu^\pm)\right]^2 \;,
\end{split} \end{equation}
where ${\rm BR}(\smu^\pm \to \chi \mu^\pm)$ is the branching ratio of the smuon into a neutralino and a muon, and $\sigma(e^+ e^- \to \smu^+\smu^-)$ is the production cross section of the $\smu$ mass eigenstate considered which is obtained via Eq.~\eqref{eq:xsecpol} after inserting and integrating the relevant differential cross sections Eq.~\eqref{eq:xsecii} or~\eqref{eq:xsec12}. As discussed earlier, in our simplified model, the smuon branching ratio is ${\rm BR}(\smu^\pm \to \chi \mu^\pm) = 1$ regardless of the values of $\kappa$ and $\alpha$ as long as $m_{\smu} > m_{\chi} + m_{\mu}$, and $0$ otherwise. However, in extended models where additional decay channels are allowed, e.g. in SUSY models with additional states lighter than the smuon, the value of the branching ratio can differ from one.

Likewise, for stau production in the ($\mu^+ \mu^- + \ME$) final state, the cross section is given by 
\begin{equation} \begin{split}
   &\sigma(e^+ e^- \to \stau^+ \stau^- \to \mu^+ \mu^- + \ME) \\
   &= \sigma(e^+ e^- \to \stau^+ \stau^-) \times \left[{\rm BR}(\stau^\pm \to \chi \tau^\pm)\right]^2 \\
   &\qquad \times \left[{\rm BR}(\tau^\pm \to \mu^\pm \nu_\tau \nu_\mu)\right]^2 \;,
\end{split} \end{equation}
where again the branching ratio ${\rm BR}(\stau^\pm \to \chi \tau^\pm) = 1$ as long as $m_{\stau} > m_\chi + m_\tau$. The production cross section for the ($e^\pm \mu^\mp + \ME$) final state is 
\begin{equation} \begin{split}
   &\sigma(e^+ e^- \to \stau^+ \stau^- \to e^\pm \mu^\mp + \ME) \\
   &= \sigma(e^+ e^- \to \stau^+ \stau^-) \times \left[{\rm BR}(\stau^\pm \to \chi \tau^\pm)\right]^2 \\
   &\qquad \times 2 \, {\rm BR}(\tau^\pm \to e^\pm \nu_\tau \nu_e) \times {\rm BR}(\tau^\pm \to \mu^\pm \nu_\tau \nu_\mu) \;.
\end{split} \end{equation}
Note that due to the branching ratios of the leptonic $\tau$ decays
\begin{equation}
   {\rm BR}(\tau^\pm \to e^\pm \nu_\tau \nu_e) \approx {\rm BR}(\tau^\pm \to \mu^\pm \nu_\tau \nu_\mu) \approx 18\,\% \;,
\end{equation}
the relevant signal cross sections for staus are a suppressed by a factor of $\sim 1/30$ compared to the smuon signal cross sections.

\section{Collider Simulations} \label{sec:BP_sim}
In order to estimate the sensitivity of future $e^+e^-$ colliders to charged lepton partners, we numerically simulate signal and background events. We use \texttt{MadGraph5\_aMC@NLO\_v2.6.1}~\cite{Alwall:2014hca} for the generation of the hard event, \texttt{pythia6}~\cite{Sjostrand:2006za} for showering\footnote{We use \texttt{pythia6} because the more recent \texttt{pythia8}~\cite{Sjostrand:2014zea} does not support photon beams in the initial state.}, and the fast detector simulation \texttt{Delphes3}~\cite{deFavereau:2013fsa}. For (charged lepton + $\ME$) final states at an $e^+e^-$ collider, the results of showering and detector effects are only to smear the momenta of the detectable final states and to account for loss of charged leptons due to geometric detector coverage and mis-identification. Since the energy reconstruction of charged leptons is comparatively easy at $e^+e^-$ colliders, the precise implementation of the detector does not have large effects on the sensitivity in (charged lepton + $\ME$) final states. For concreteness we use the \texttt{Delphes} detector card from Ref.~\cite{Potter:2016pgp} based on proposed ILC detector designs.\footnote{We note that more sophisticated simulation chains are available for future lepton colliders (see for example Refs.~\cite{Baer:2019gvu, Habermehl:2020njb, Zarnecki:2020swm}). However, the chain we have chosen is sufficient for the analysis here, see for example the discussion in Ref.~\cite{Potter:2020kfv}.}

We simulate background and signal processes for two center-of-mass energies, $\sqrt{s} = 250\,$GeV and $\sqrt{s} = 500\,$GeV, as well as for three different polarization settings, $P = \left(0,0\right)$, $P = \left(0.8,0.3\right)$, and $P = \left(-0.8,-0.3\right)$, as defined in Eq.~\eqref{eq:PRdef}.

We describe the background and signal simulations for the ($\mu^+\mu^- + \ME$) and ($e^\pm\mu^\mp + \ME$) final states in Secs.~\ref{sec:BP_smuon} and~\ref{sec:BP_stau}, respectively. In order to define a set of kinematic cuts for the event selection, we compare the kinematic distributions of the backgrounds to those of the signal for benchmark values of the slepton and neutralino masses. We discuss the most relevant features of the kinematic cuts employed here in the respective subsections; a more detailed description can be found in the Appendices~\ref{app:smuon} and~\ref{app:stau}. 

\begin{figure*}
   \includegraphics[width=0.49\linewidth]{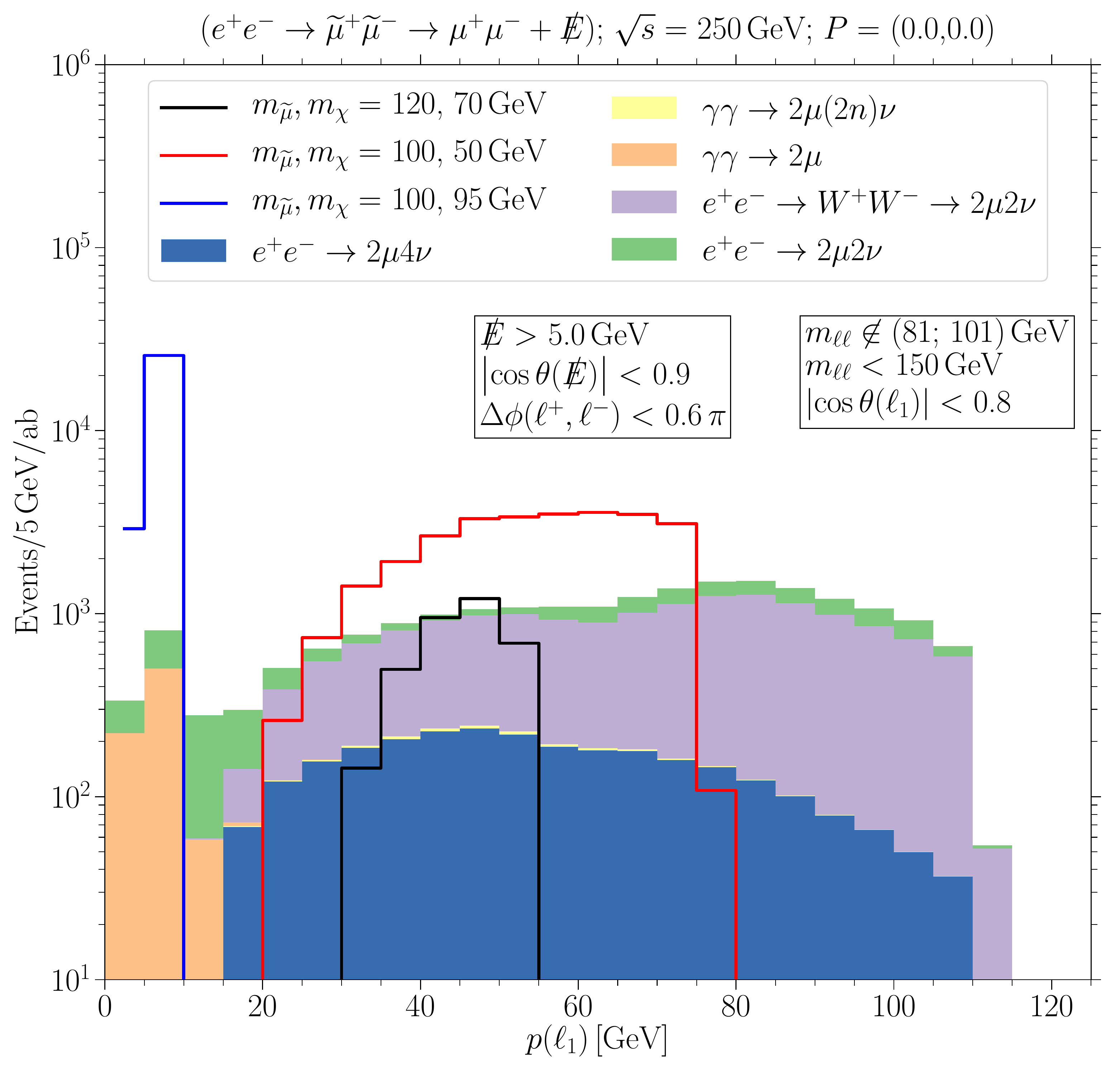}
   \includegraphics[width=0.49\linewidth]{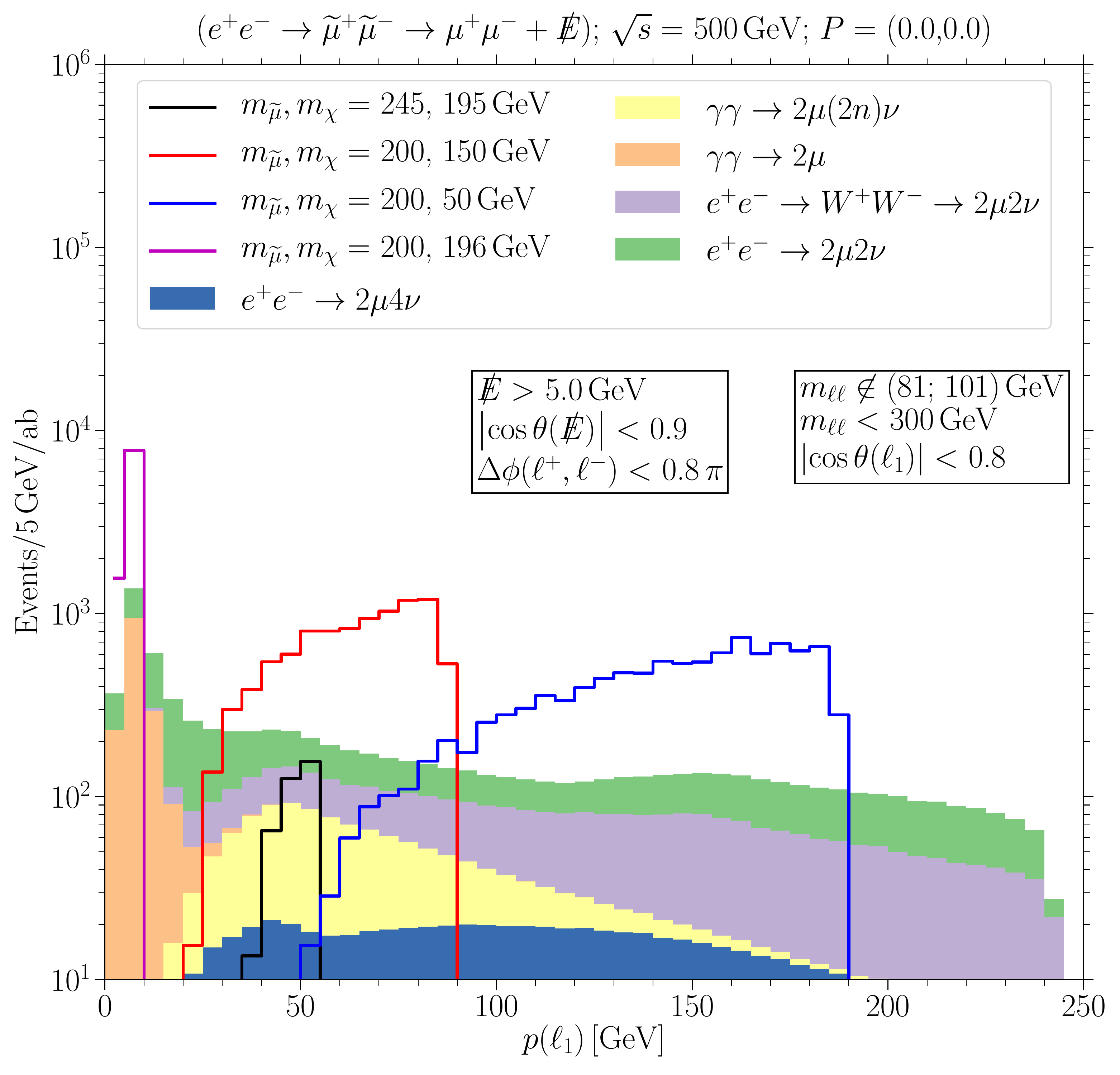}
   \caption{Distribution of leading lepton momenta $p(\ell_1)$ in the ($\mu^+\mu^- + \ME$) final state after kinematic cuts described in the text and denoted in the figures. The left panel is for a center-of-mass energy of $\sqrt{s} = 250\,$GeV and the right panel for $\sqrt{s} = 500\,$GeV. Both panels are for unpolarized beams $P = (0.0,0.0)$. In each panel, the stacked histogram shows the different background components as indicated in the legend. Note that the ($e^+e^- \to 2\mu2\nu$) background category (green) excludes events which stem from on-shell $W^+W^-$ pair production, such events are shown separately (purple). The ($\gamma\gamma \to 2\mu(2n)\nu$) background category includes both ($\gamma\gamma \to 2\mu2\nu$) and ($\gamma\gamma \to 2\mu4\nu$) events. The solid lines show ($e^+ e^- \to \gamma / Z \to \smu^+ \smu^- \to \mu^+ \mu^- \chi \chi$) signal distributions for the various benchmark values of the smuon mass $m_{\smu}$ and neutralino mass $m_\chi$ indicated in the legend.}
   \label{fig:smuons}
\end{figure*}

\subsection{($\mu^+\mu^- + \ME$) Final State} \label{sec:BP_smuon}

A number of SM processes can give rise to ($\mu^+\mu^- + \ME$) final states at an $e^+e^-$ collider. We simulate SM processes giving rise to a pair of opposite charge muons and zero, two, or four neutrinos (of any flavor). For each final state, we simulate event samples stemming from $e^+ e^-$ collisions as well as from $\gamma\gamma$ collisions. As discussed above, the latter initial state arises from the photon components of the electron and positron beams, which become non-negligible at the center-of-mass energies considered here. We use the default implementation of the electron's photon component in \texttt{MadGraph5}, which is based on the parameterization in Ref.~\cite{Frixione:1993yw}.

\begin{table}
   \setlength{\extrarowheight}{2pt}
   \begin{tabular}{L{4cm} C{2cm}}
      \hline\hline
      Background process & \# of samples \\
      \hline
      $e^+e^- \to \mu^+\mu^-$ & $10^7$ \\
      $e^+e^- \to \mu^+\mu^- + 2\nu$ & $10^7$ \\
      $e^+e^- \to \mu^+\mu^- + 4\nu$ & $5 \times 10^6$\\
      $\phantom{e^+} \gamma\gamma \to \mu^+\mu^-$ & $10^9$\\
      $\phantom{e^+} \gamma\gamma \to \mu^+\mu^- + 2\nu$ & $10^7$ \\
      $\phantom{e^+} \gamma\gamma \to \mu^+\mu^- + 4\nu$ & $10^7$ \\
      \hline \hline
   \end{tabular}
   \caption{Number of events produced for each of the different background categories for the ($\mu^+\mu^- + \ME$) final state.}
   \label{tab:Nbkg}
\end{table}

\begin{table}
   \setlength{\extrarowheight}{2pt}
   \begin{tabular}{C{2.8cm} C{2.7cm} C{2.7cm}}
      \hline\hline
      Kinematic Variable & $\sqrt{s} = 250\,$GeV & $\sqrt{s} = 500\,$GeV \\
      \hline
      $\ME$ & $> 5\,$GeV & $> 5\,$GeV \\
      $\left|\cos\theta(\ME)\right|$ & $< 0.9$ & $< 0.9$ \\
      $\Delta\phi(\ell^+,\ell^-)$ & $< 0.6\,\pi$ & $< 0.8\,\pi$ \\
      \hline
      $m_{\ell\ell}$ & $\not\in \left(81; 101\right)\,$GeV & $\not\in \left(81; 101\right)\,$GeV \\
      $m_{\ell\ell}$ & $< 150\,$GeV & $< 300\,$GeV \\
      $\left|\cos\theta(\ell_1)\right|$ & $< 0.8$ & $< 0.8$ \\
      \hline \hline
   \end{tabular}
   \caption{Kinematic cuts for event selection for smuon searches in the ($\mu^+ \mu^- + \ME$) final state.}
   \label{tab:cuts_smu}
\end{table}

\begin{figure}
   \includegraphics[width=\linewidth]{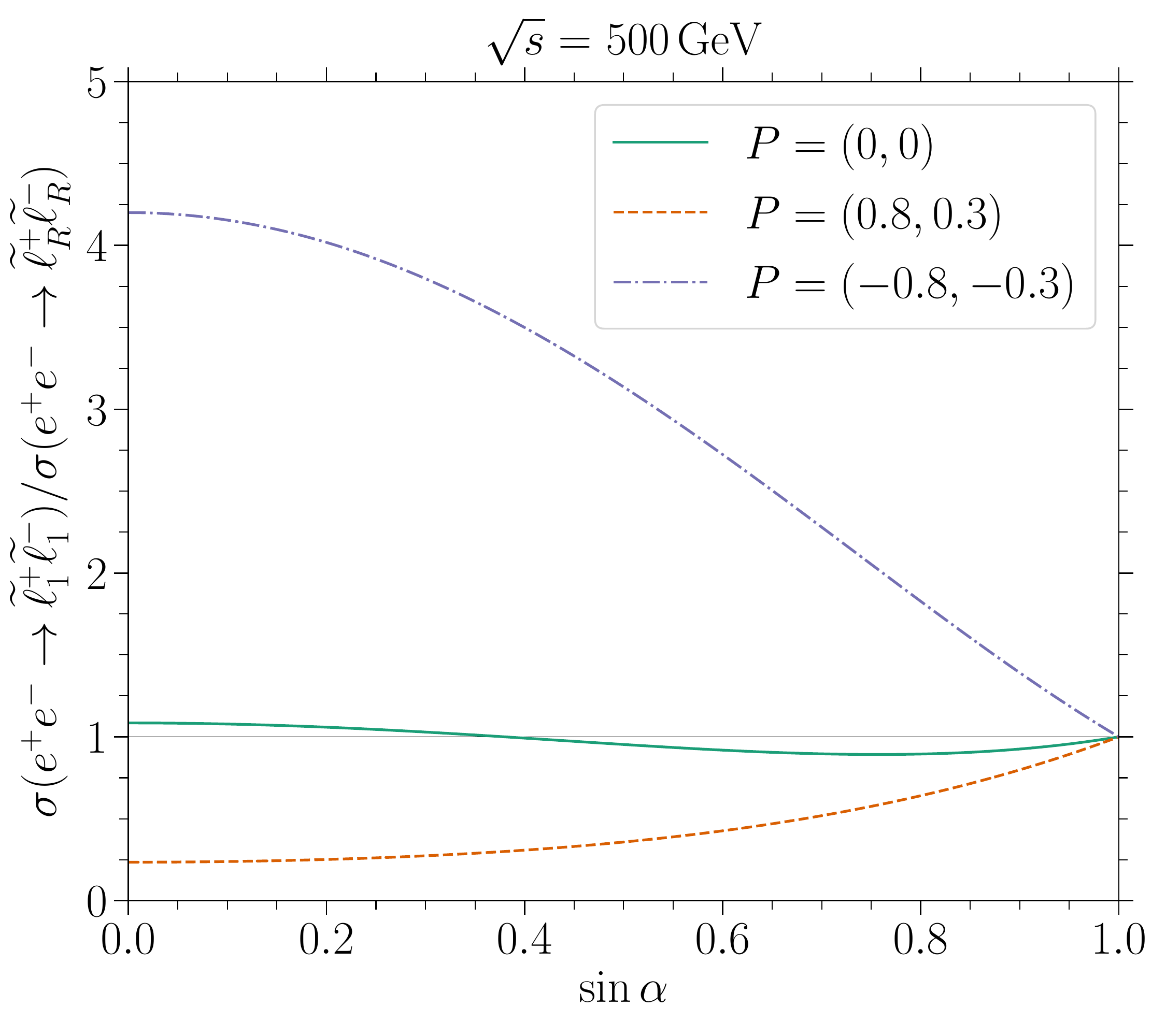}
   \caption{Scaling of the slepton production cross section with the left-right mixing angle $\sin\alpha$ for different beam polarizations as indicated in the legend. $\sin\alpha=0$ corresponds to left-handed sleptons, $\sell_1 = \sell_L$, and $\sin\alpha=1$ to right-handed sleptons, $\sell_1 = \sell_R$.}
   \label{fig:lr_scaling}
\end{figure}

Out of these background processes, ($e^+ e^- \to \mu^+\mu^-$) events have no $\ME$ at the truth level. Nonetheless, such events can feature $\ME > 0$ in the detector because of mis-measured momenta of the leptons. ($\gamma\gamma \to \mu^+\mu^- + \ME$) events on the other hand have non-vanishing $\ME$ in the detector frame already at the truth level, because the center-of-mass frame of the $\gamma\gamma$ system can have sizable boosts along the beam axis with respect to the detector frame. Note though that ($\gamma\gamma \to \mu^+\mu^- + \ME$) events have no true transverse missing energy. All other background processes considered here feature true missing transverse energy from the neutrinos. In Tab.~\ref{tab:Nbkg} we list the number of events simulated for each of the background categories, which were chosen to yield sufficiently well-sampled background distributions after kinematic cuts. Note that we oversample ($\gamma\gamma \to \mu^+\mu^-$) events because only the tails of the distributions from such events will survive our kinematic cuts.

In order to differentiate the signal from the background spectra, we employ a set of kinematic observables:
\begin{itemize}
   \item the momentum of the leading (charged) lepton, $p(\ell_1)$,
   \item the (cosine of the) polar angle of the leading lepton with respect to the beam axis, $\cos\theta(\ell_1)$,
   \item the invariant mass of the charged lepton system, $m_{\ell\ell}$,
   \item the opening angle between the charged leptons in the plane transverse to the beam axis, $\Delta\phi(\ell^+,\ell^-)$,
   \item the missing energy, $\ME$,
   \item the (cosine of the) polar angle of $\ME$ with respect to the beam axis, $\cos\theta(\ME)$.
\end{itemize}

We employ a series of kinematic cuts on these variables in order to suppress background events, see Appendix~\ref{app:smuon} for a more detailed discussion. The numerical values of the cuts can be found in Tab.~\ref{tab:cuts_smu} and the labels of Fig.~\ref{fig:smuons}. 

In order to reduce the backgrounds from ($e^+e^- \to \mu^+ \mu^-$) and ($\gamma\gamma \to \mu^+ \mu^-$) events, we employ a lower cut on $\ME$, an upper cut on $\left|\cos\theta(\ME)\right|$, and an upper cut on $\Delta\phi(\ell^+,\ell^-)$. These cuts effectively demand events to feature transverse missing energy. 

In order to reduce the remaining backgrounds, we employ two separate cuts on $m_{\ell\ell}$: Background events where the charged leptons stem from the decay of an on-shell $Z$-boson are suppressed by excluding events where $m_{\ell\ell} \simeq m_Z$. Furthermore, in the ($e^+e^- \to \smu^+\smu^- \to \mu^+\mu^-\chi\chi$) signal events, some of the center-of-mass energy is absorbed by the production of the (stable) massive neutralinos in the final state. For the background events on the other hand, no heavy particles are present in the final state. Thus, we utilize an upper cut on $m_{\ell\ell}$ to reject backgrounds where all neutrinos are soft and hence $m_{\ell\ell} \to \sqrt{s}$. Finally, signal events are produced via $s$-channel processes, while much of the background stems from $t$-channel processes, e.g. ($e^+ e^- \to W^+W^-$) pair production via $t$-channel exchange of a neutrino. Compared to $s$-channel processes, $t$-channel processes result in charged leptons that are more collimated along the beam axis, thus, we use an upper cut on $|\cos\theta(\ell_1)|$.

In Fig.~\ref{fig:smuons} we show the distributions of leading lepton momenta, $p(\ell_1)$, after these cuts for the backgrounds (stacked histograms) together with the signal distributions for various benchmark values of the smuon and neutralino masses. The $p(\ell_1)$ distribution is the final discriminating variable; in Sec.~\ref{sec:sens} we project the sensitivity of future $e^+e^-$ colliders as a function of smuon and neutralino masses by optimizing a signal window in $p(\ell_1)$ in addition to the kinematic cuts described above. Note that here and in all other figures showing background spectra, we select (post simulation) background events in the $(\mu^+\mu^- + 2\nu)$ final state which feature a pair of intermediate on-shell $W^+W^-$ bosons and show them separately from the other $(\mu^+\mu^- + 2\nu)$ background. 

The shape of the $p(\ell_1)$ spectrum is controlled by $m_{\smu}$ and the ratio $m_\chi/m_{\smu}$. It is straightforward to work out from the kinematics that the (true) momenta of the muons lie in between~\cite{Battaglia:2005zf}
\begin{equation}
   p(\mu^\pm)_{\substack{{\rm max} \\ {\rm min} }} = \frac{m_{\smu}}{2} \left(1-\frac{m_\chi^2}{m_{\smu}^2}\right) \gamma \left(1 \pm \beta \right)\;,
\end{equation}
where we neglected the muon mass, $\beta = \sqrt{1-4m_{\smu}^2/s}$ is the boost of the smuon in the detector frame, and $\gamma=1/\sqrt{1-\beta^2}$. For compressed spectra, $m_\chi \to m_{\smu}$, the muons become soft, even for $m_{\smu} \ll \sqrt{s}/2$. As we will see in Sec.~\ref{sec:sens}, $e^+e^-$ colliders remain sensitive to slepton pair production even for $m_\chi \to m_{\smu}$. 

\begin{figure*}
   \includegraphics[width=0.49\linewidth]{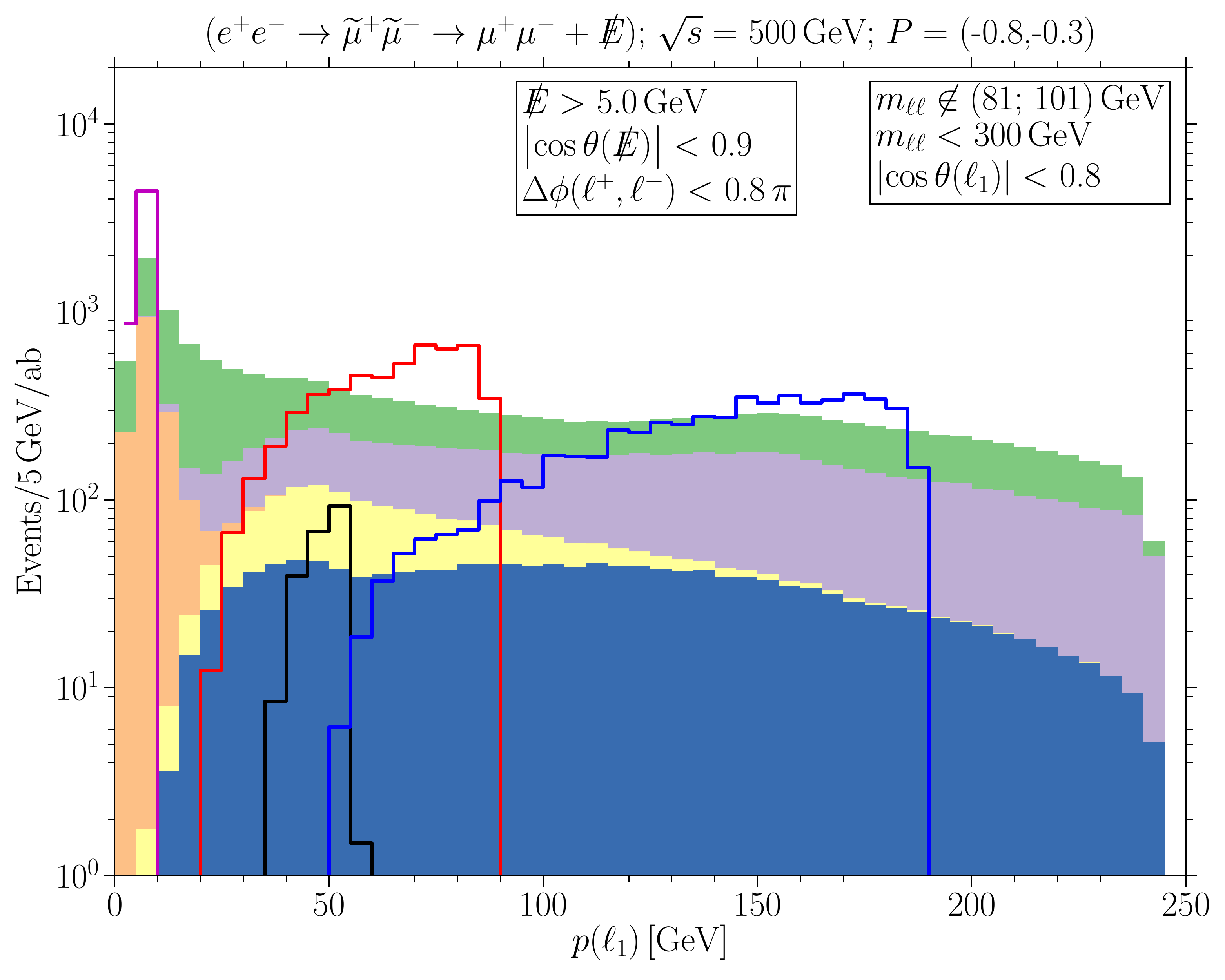}
   \includegraphics[width=0.49\linewidth]{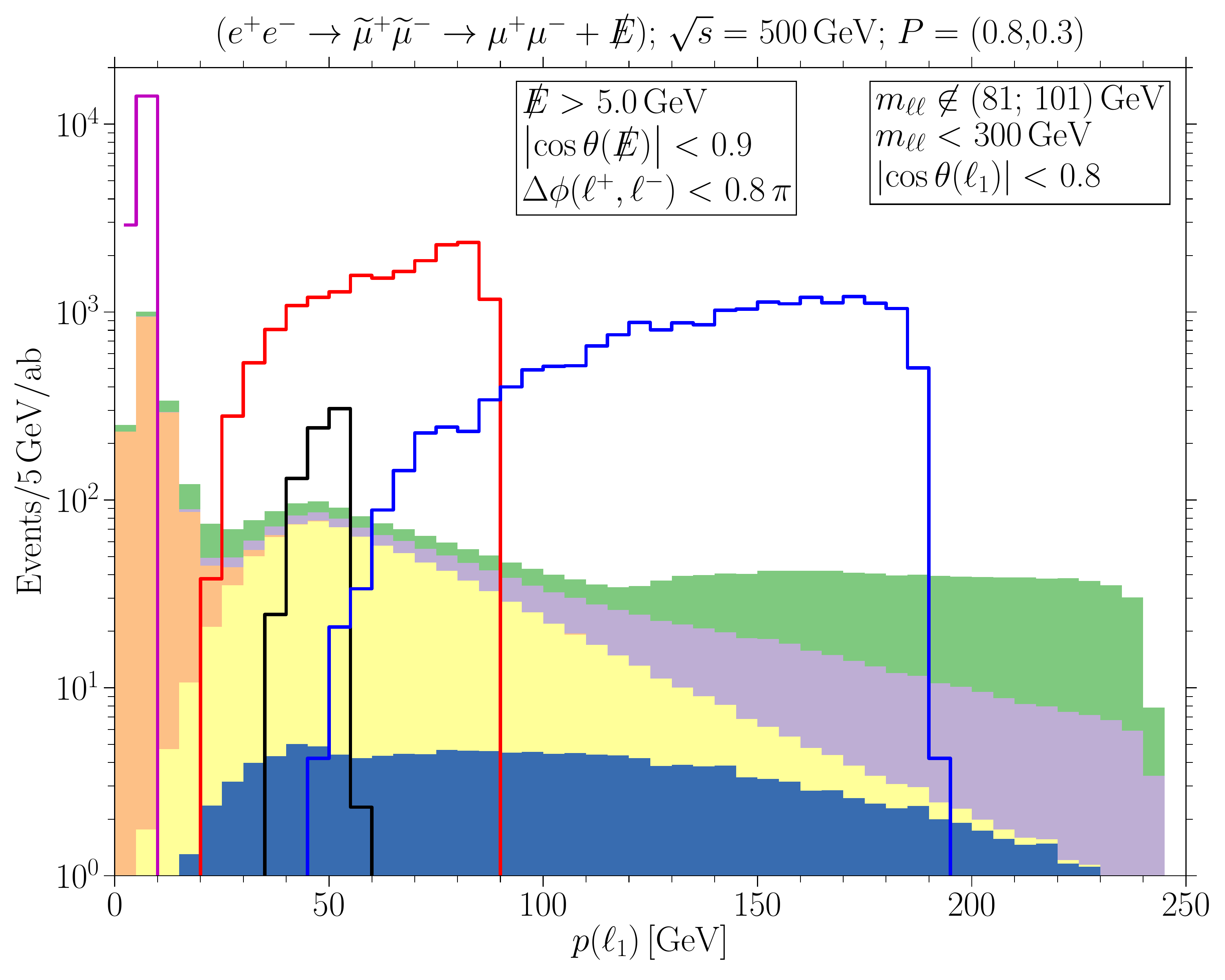}
   \caption{Same as the right panel of Fig.~\ref{fig:smuons} but for mostly left-handed $P=(-0.8,-0.3)$ and mostly right-handed $P=(0.8,0.3)$ polarization configurations of the electron/positron beams. We remind the reader that we express polarizations via $P = \left[ P_R(e^-), P_R(e^+) \right]$, see Eq.~\eqref{eq:PRdef} for the definition of $P_R$. The colors for the respective background components and lines of the benchmark signal distributions are the same as in the right panel of Fig.~\ref{fig:smuons}. Note that the relative normalization of the signal distributions compared to that of the background is largely insensitive to the left-right mixing of the sleptons when the electron/positron beams are unpolarized. The relative normalization of the signal to background varies the most for the production of right-handed sleptons in polarized beam configurations, as shown here and discussed in the text.}
   \label{fig:smuons_pol}
\end{figure*}

The signal spectra in Fig.~\ref{fig:smuons} are for production of right-handed smuons $\smu_R$. As discussed in Sec.~\ref{sec:model}, considering instead left-handed smuons $\smu_L$ or left-right mixed mass eigenstates ($e^+ e^- \to \smu_1 \smu_1 \to \mu^+ \mu^- \chi \chi$) with $\smu_1 = \smu_L \cos\alpha + \smu_R \sin\alpha$, would only have changed the production cross section, i.e. the normalization, but not the shape of the kinematic distributions. In Fig.~\ref{fig:lr_scaling} we show the production cross sections for smuons and staus as a function of the left-right mixing angle $\alpha$. For non-polarized beams with $P = (0,0)$, the production cross section is only weakly dependent on $\alpha$. For mostly right-handed [left-handed] initial electron and positron beams, $P = (0.8, 0.3)$ [$P = (-0.8,-0.3)$] on the other hand, the production cross section for left-handed smuons/staus is a factor of $\sim 4$ smaller [larger] than the production cross section for right-handed smuons/staus.

Similarly, much of the background involves the production of electroweak gauge bosons. Since $W$-bosons couple to pairs of left-handed fermions, but not to right-handed fermions, the normalization of various background components also depends on the beam polarization. Generally, the more right-handed the electron and positron beam polarizations become, the less SM background is produced in the ($\mu^+\mu^- + \ME$) final state. In Fig.~\ref{fig:smuons_pol} we show the distribution of the leading lepton momenta for the backgrounds together with benchmark signal spectra for right-handed smuons for mostly left-handed beam polarizations $P = \left(-0.8,-0.3\right)$ in the left panel and mostly right-handed beam polarization $P = \left(0.8,0.3\right)$ in the right panel. As anticipated, mostly right-handed beam polarization is the preferred configuration when probing right-handed smuons. The sensitivity to left-handed smuons on the other hand is virtually independent of the beam polarization since the signal production cross section and those of the most relevant backgrounds scale similarly with the beam polarizations.

Let us briefly comment on probing stau pair production in the ($\mu^+\mu^- + \ME$) final state. As discussed in Sec.~\ref{sec:model}, the signal cross section is suppressed by a factor of $\left[{\rm BR}(\tau^\pm \to \mu^\pm \nu_\tau \nu_\mu)\right]^2 \approx 1/30$ compared to the smuon case. Furthermore, because of the production of the additional neutrinos in the tau decays, the muons in the final state will be softer than in the smuon case. In combination, these effects make it rather challenging to probe stau pair production in the ($\mu^+\mu^- + \ME$) final state, even when employing more aggressive kinematic cuts (see Fig.~\ref{fig:stau_2mu} and associated discussion in Appendix~\ref{app:smuon}). The ($e^\pm\mu
^\mp + \ME$) final state discussed in the subsequent subsection is a more promising alternative for probing staus.

\begin{figure*}
   \includegraphics[width=.49\linewidth]{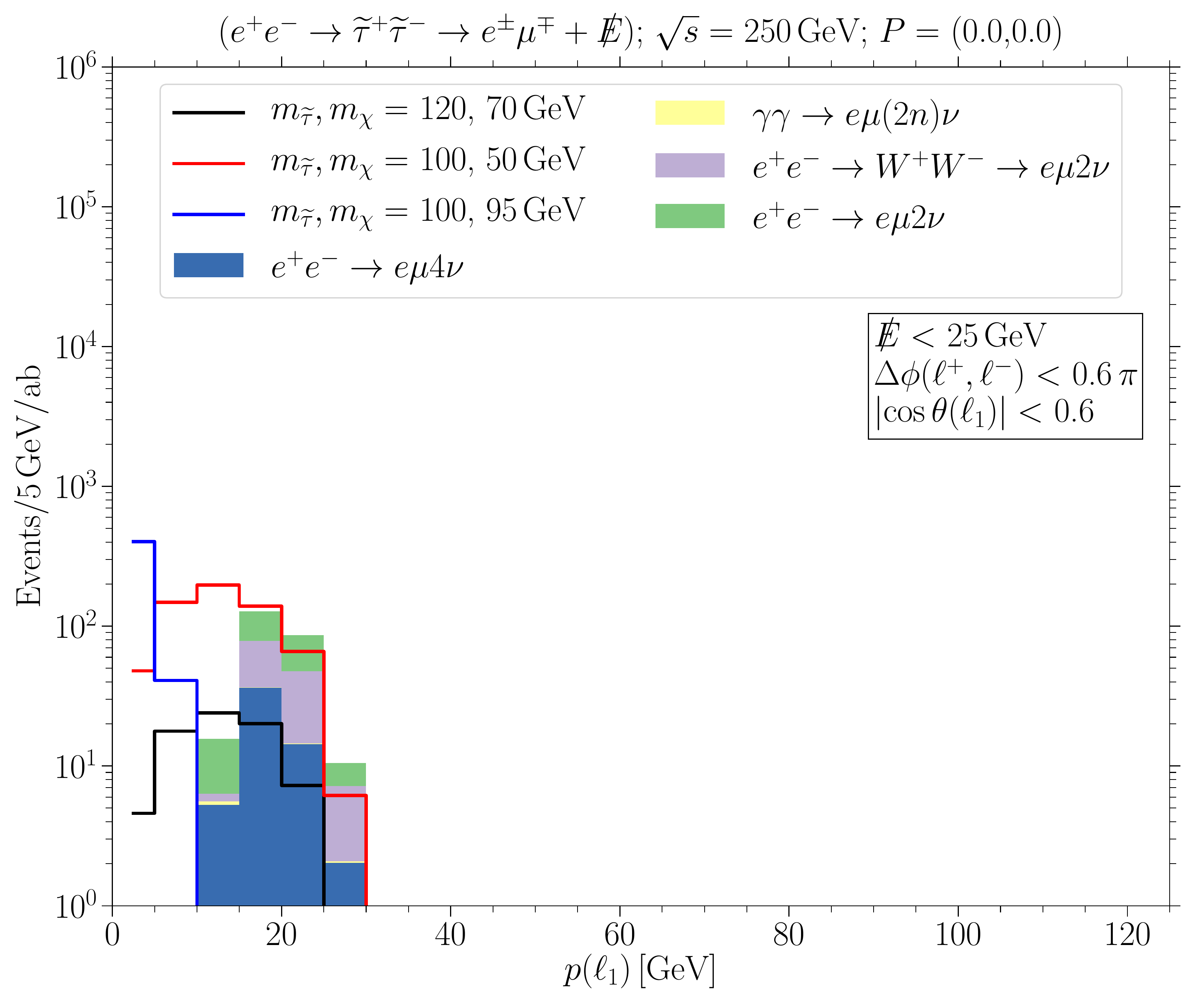}
   \includegraphics[width=.49\linewidth]{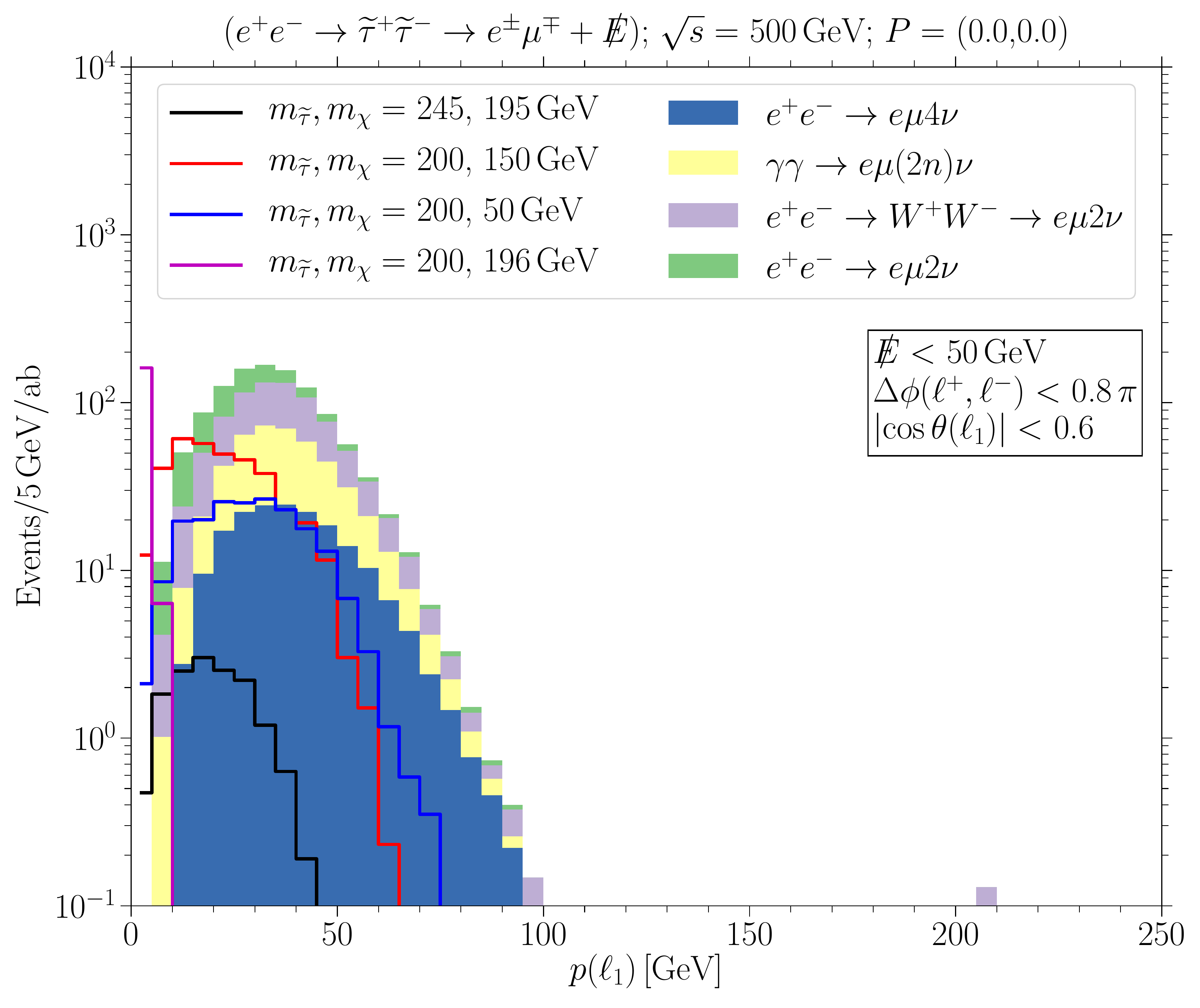}
   \caption{Distribution of leading lepton momenta $p(\ell_1)$ in the ($e^\pm\mu^\mp + \ME$) final state after kinematic cuts described in the text and denoted in the figures. The left panel is for a center-of-mass energy of $\sqrt{s} = 250\,$GeV and the right panel for $\sqrt{s} = 500\,$GeV, and both panels are for unpolarized beams $P = (0.0,0.0)$. In each panel, the stacked histogram shows the different background components as indicated in the legend. Note that the ($e^+e^- \to e\mu2\nu$) background category (green) excludes events which stem from on-shell $W^+W^-$ pair production, such events are shown separately (purple). The ($\gamma\gamma \to e\mu(2n)\nu$) background category includes both ($\gamma\gamma \to e\mu2\nu$) and ($\gamma\gamma \to e\mu4\nu$) events. The solid lines show ($e^+ e^- \to \gamma / Z \to \stau^+ \stau^- \to e^\pm \mu^\mp \chi \chi \nu_e \nu_\mu \nu_\tau \bar{\nu}_\tau$) signal distributions for the various benchmark values of the stau mass $m_{\stau}$ and neutralino mass $m_\chi$ indicated in the legend.}
   \label{fig:stau}
\end{figure*}

\begin{table}
   \setlength{\extrarowheight}{2pt}
   \begin{tabular}{C{2.8cm} C{2.7cm} C{2.7cm}}
      \hline\hline
      Kinematic Variable & $\sqrt{s} = 250\,$GeV & $\sqrt{s} = 500\,$GeV \\
      \hline
      $\ME$ & $> 5\,$GeV & $> 10\,$GeV \\
      $\left|\cos\theta(\ME)\right|$ & $< 0.6$ & $< 0.6$ \\
      $\Delta\phi(\ell^+,\ell^-)$ & $< 0.6\,\pi$ & $< 0.6\,\pi$ \\
      \hline
      $m_{\ell\ell}$ & $\not\in \left(81; 101\right)\,$GeV & $\not\in \left(81; 101\right)\,$GeV \\
      $m_{\ell\ell}$ & $< 75\,$GeV & $< 150\,$GeV \\
      $\left|\cos\theta(\ell_1)\right|$ & $< 0.8$ & $< 0.8$ \\
      \hline \hline
   \end{tabular}
   \caption{Kinematic cuts for event selection for stau searches in the ($\mu^+ \mu^- + \ME$) final state.}
   \label{tab:cuts_stau_2mu}
\end{table} 

\subsection{($e^\pm\mu^\mp + \ME$) Final State} \label{sec:BP_stau}

\begin{table}
   \setlength{\extrarowheight}{2pt}
   \begin{tabular}{C{2.8cm} C{2.7cm} C{2.7cm}}
      \hline\hline
      Kinematic Variable & $\sqrt{s} = 250\,$GeV & $\sqrt{s} = 500\,$GeV \\
      \hline
      $\ME$ & $<25\,$GeV & $< 50\,$GeV \\
      $\Delta\phi(\ell^+,\ell^-)$ & $< 0.6\,\pi$ & $< 0.8\,\pi$ \\
      $\left|\cos\theta(\ell_1)\right|$ & $< 0.6$ & $< 0.6$ \\
      \hline \hline
   \end{tabular}
   \caption{Kinematic cuts for event selection for stau searches in the ($e^\pm \mu^\mp + \ME$) final state.}
   \label{tab:cuts_stau_emu}
\end{table}

The ($e^\pm \mu^\mp + \ME$) final state is in many respects similar to the ($\mu^+\mu^- + \ME$) final state discussed in the previous subsection. However, some SM processes which give rise to backgrounds in the ($\mu^+\mu^- + \ME$) final state are absent for ($e^\pm \mu^\mp + \ME$), which affects our choice of kinematic cuts. In particular, ($e^+ e^-/\gamma\gamma \to e^\pm \mu^\mp$) processes are forbidden by flavor conservation; any final state at $e^+e^-$ colliders containing exactly one electron and one muon must contain at least two neutrinos. Thus, there is no need to use the cuts on $\ME$ and $\left|\cos\theta(\ME)\right|$ we employed in the ($\mu^+\mu^- + \ME$) final state to reduce the ($e^+ e^-/\gamma\gamma \to \mu^+ \mu^-$) events which have no true (transverse) missing energy. Furthermore, for the ($\mu^+\mu^- + \ME$) final state, we excluded events where the invariant mass of the charged lepton system satisfied $m_{\ell\ell} \simeq m_Z$ to suppress backgrounds where the charged lepton pair stems from the decay of an (on-shell) $Z$-boson. Since ($Z \to e^\pm \mu^\mp$) decays are forbidden, the corresponding background is absent in the ($e^\pm \mu^\mp + \ME$) final state and we do not use such a cut.

On the other hand, as already mentioned at the end of Sec.~\ref{sec:BP_smuon}, the charged leptons arising from stau pair production are softer than in the smuon case because of the additional neutrinos produced in the ($\tau^\pm \to \ell^\pm \nu_\tau \nu_\ell$) decays. Furthermore, the momenta of the four neutrinos and two neutralinos in the final state will partially cancel. Hence, signal events typically feature relatively small $\ME$, and we use an upper cut on $\ME$. Because of the massive neutralinos in the final state, signal events on average feature smaller opening angles between the charged leptons $\Delta\phi(\ell^+,\ell^-)$ than the background events for which the final state contains only massless neutrinos in addition to the charged leptons. Thus, we use an upper cut on $\Delta\phi(\ell^+,\ell^-)$. Finally, as discussed above for the ($\mu^+\mu^- + \ME$) final state, signal events are produced via $s$-channel processes while much of the background arises from $t$-channel processes. Thus, we use an upper cut on $\left|\cos\theta(\ell_1)\right|$ to select events. 

In Fig.~\ref{fig:stau} we show the distributions of the leading lepton momenta $p(\ell_1)$ of the backgrounds compared to the signal distributions for benchmark values of the stau and neutralino masses after these kinematic cuts. We simulate the same number of background events as for the analogous background in the ($\mu^+\mu^- + \ME$) final state, cf. Tab.~\ref{tab:Nbkg}. Note that the leading lepton $\ell_1$ can either be the electron or muon in the final state. In all signal and background processes, electrons and muons can be interchanged (when interchanging the neutrino flavors accordingly). Furthermore, muons and electrons have masses much smaller than the center-of-mass energy of the collisions considered here. Thus, treating the kinematic variables for the muon and electron in the final state separately would not yield additional information.

Comparing Fig.~\ref{fig:stau} with Fig.~\ref{fig:smuons}, we can see the impact of the differences of both the background and signal spectra discussed above between stau searches in the ($e^\pm \mu^\pm + \ME$) final state and smuon searches in the ($\mu^+ \mu^- + \ME$) final state. Since there is no analogous process to ($\gamma\gamma \to \mu^+ \mu^-$) in the ($e^\pm \mu^\mp + \ME$) final state, there is much less background at small values of the leading lepton momentum. This is advantageous for stau searches, because for the reasons discussed above, the charged leptons from staus are softer than those arising from smuon pair production. The effect of changing the polarization of the electron/positron beams on the signal and background spectra is analogous to the smuons in the ($\mu^+ \mu^- + \ME$) final state (see the discussion in Sec.~\ref{sec:BP_smuon}). 

\section{Projected Sensitivity} \label{sec:sens}

\begin{figure*}
   \includegraphics[width=0.49\linewidth]{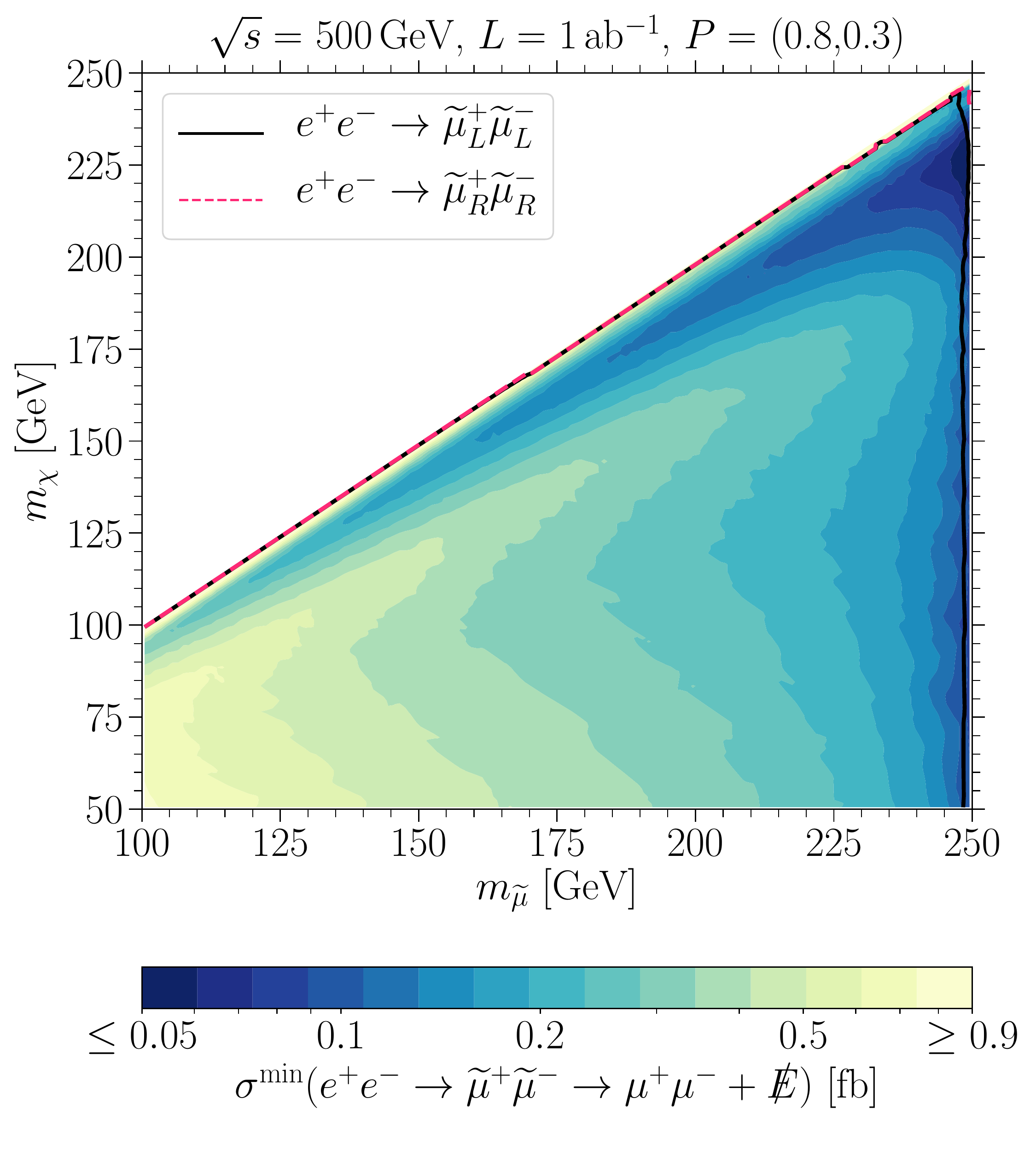}
   \includegraphics[width=0.49\linewidth]{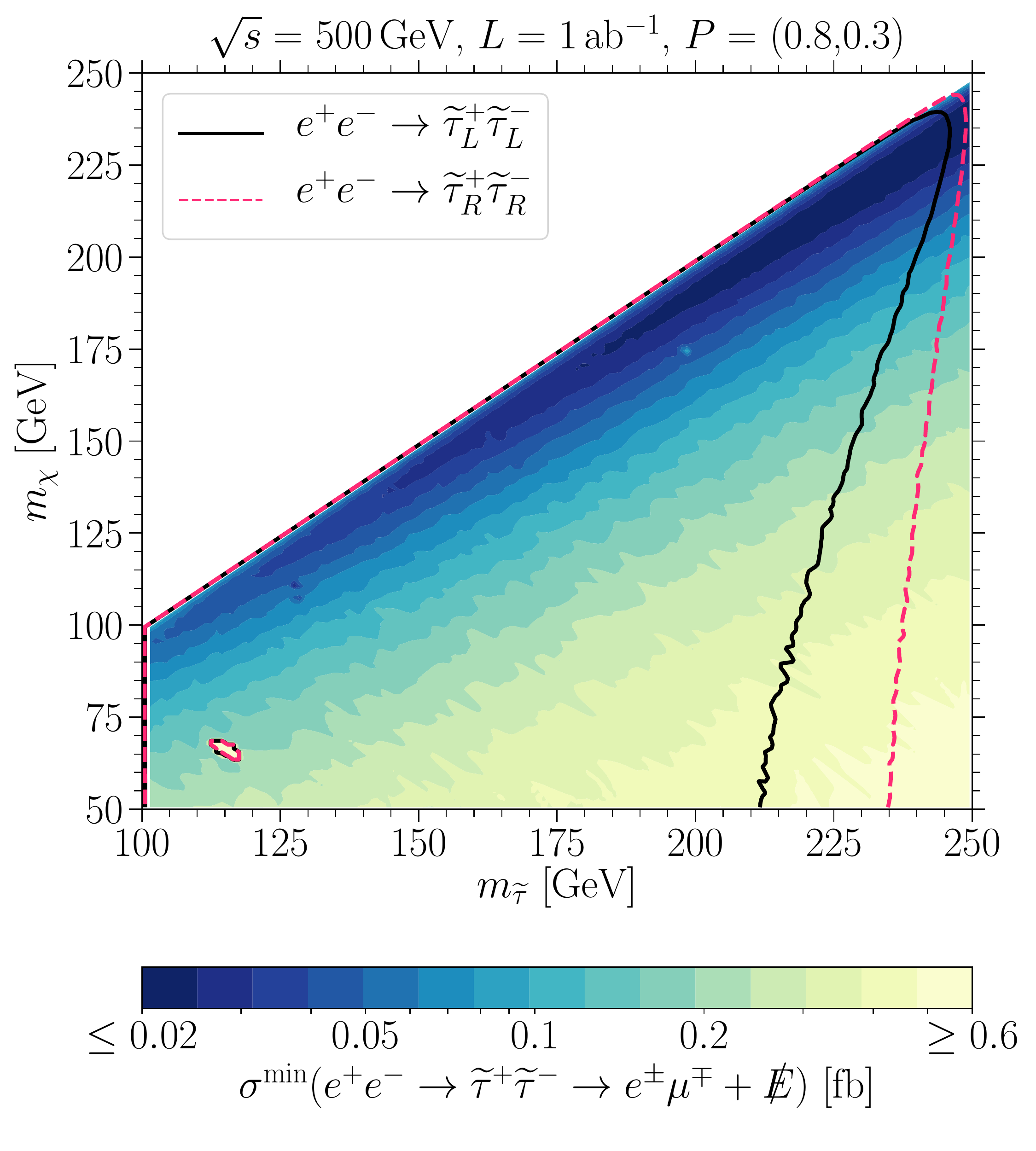}
   \caption{The left [right] panel shows our projections for the smallest smuon [stau] signal cross section $\sigma^{\rm min}$ which could be probed at an $\sqrt{s}=500\,$GeV $e^+e^-$ collider with $L=1\,{\rm ab}^{-1}$ of data and mostly right-handed beams $P = \left( 0.8, 0.3\right)$ in the ($\mu^+\mu^- + \ME$) [($e^\pm \mu^\mp + \ME$)] final state. For reference, the solid black (dashed red) line indicates where $\sigma^{\rm min}$ coincides with the cross section in the model presented in Sec.~\ref{sec:model} for left-handed (right-handed) smuons/staus. The region to the lower left of the respective contours, i.e. almost all of the parameter space where $m_{\sell} < \sqrt{s}/2$, is within the projected sensitivity. Note that in both panels, the solid black and dashed red lines overlap for $m_\chi \to m_{\sell}$. For the smuon case (left panel), the dashed red line for right-handed smuons coincides with $m_{\smu} \to 250\,$GeV for $m_\chi \lesssim 240\,$GeV and is not visible on the plot. We would like to remind the reader that the signal cross section in our model matches that in SUSY models if the respective slepton and neutralino are the next-to-lightest and lightest supersymmetric state, respectively. Note that the range of the color scale indicating the value of $\sigma^{\rm min}$ differs between the panels. The ``island'' in the right panel at $m_{\stau} \approx 115\,$GeV, $m_\chi \approx 65\,$GeV is an artifact of the statistical fluctuations in the Monte Carlo simulations of the signal (and to a lesser extent, the backgrounds).}
   \label{fig:reach_500}
\end{figure*}

In order to project the sensitivity of future $e^+e^-$ colliders to slepton production, we simulate signal samples for a grid in the masses of the slepton and the neutralino. We perform such simulations for smuons in the ($\mu^+\mu^- + \ME$) final state and for staus in both the ($\mu^+\mu^- + \ME$) and ($e^\pm \mu^\mp + \ME$) final states. We use the background simulations for the respective final states and kinematic cuts discussed in Sec.~\ref{sec:BP_sim} (see Tabs.~\ref{tab:cuts_smu}, \ref{tab:cuts_stau_2mu}, and~\ref{tab:cuts_stau_emu} for summaries of the cuts, or Figs.~\ref{fig:smuons}, \ref{fig:stau}, and~\ref{fig:stau_2mu}).

In addition to these kinematic cuts, we optimize a window in the leading lepton momentum $p(\ell_1)$ for each mass point. In order to evaluate the sensitivity, we compare the number of signal events $S$ in the signal window to the uncertainty of the number of background events $B$ via the {\it Signal to Noise Ratio} (SNR),
\begin{equation}
   {\rm SNR} = \frac{S}{\sqrt{B + \Delta^2 B^2}} \;.
\end{equation}
The denominator is the quadratic sum of the statistical error ($\sqrt{B}$) and the systematic error ($\Delta B$), which we parameterize via the relative systematic error $\Delta$. We consider a mass point the be within reach if the SNR satisfies ${\rm SNR} > 3$. In addition, we also demand the number of signal events to be $S > 5$. We assume a relative systematic uncertainty of the backgrounds of $\Delta = 5\,\%$.

In Figs.~\ref{fig:reach_500}-\ref{fig:reach_250} we present results in the $(m_{\sell},m_\chi)$ plane for center-of-mass energies $\sqrt{s} = 500\,$GeV and $\sqrt{s} = 250\,$GeV. We present results for different configurations of the longitudinal beam polarizations, and use two figures of merit: 1) we fix the integrated luminosity to $L=1\,{\rm ab}^{-1}$ ($L = 0.5\,{\rm ab}^{-1}$) for $\sqrt{s} = 500\,$GeV ($\sqrt{s} = 250\,$GeV) and compute the minimal value of the signal cross section $\sigma^{\rm min}$ which could be probed at a future collider. 2) We fix the signal cross section to the value expected in our model presented in Sec.~\ref{sec:model}, and compute the smallest integrated luminosity $L^{\rm min}$ required to probe a point of given slepton and neutralino mass. Note that the signal cross section in our model coincides with that of SUSY models such as the MSSM as long as the sought-after slepton and neutralino are the next-to-lightest and lightest supersymmetric particle, respectively. 

\subsection{$\sqrt{s}= 500\,{\rm GeV}$}

\begin{figure*}
   \includegraphics[width=0.49\linewidth]{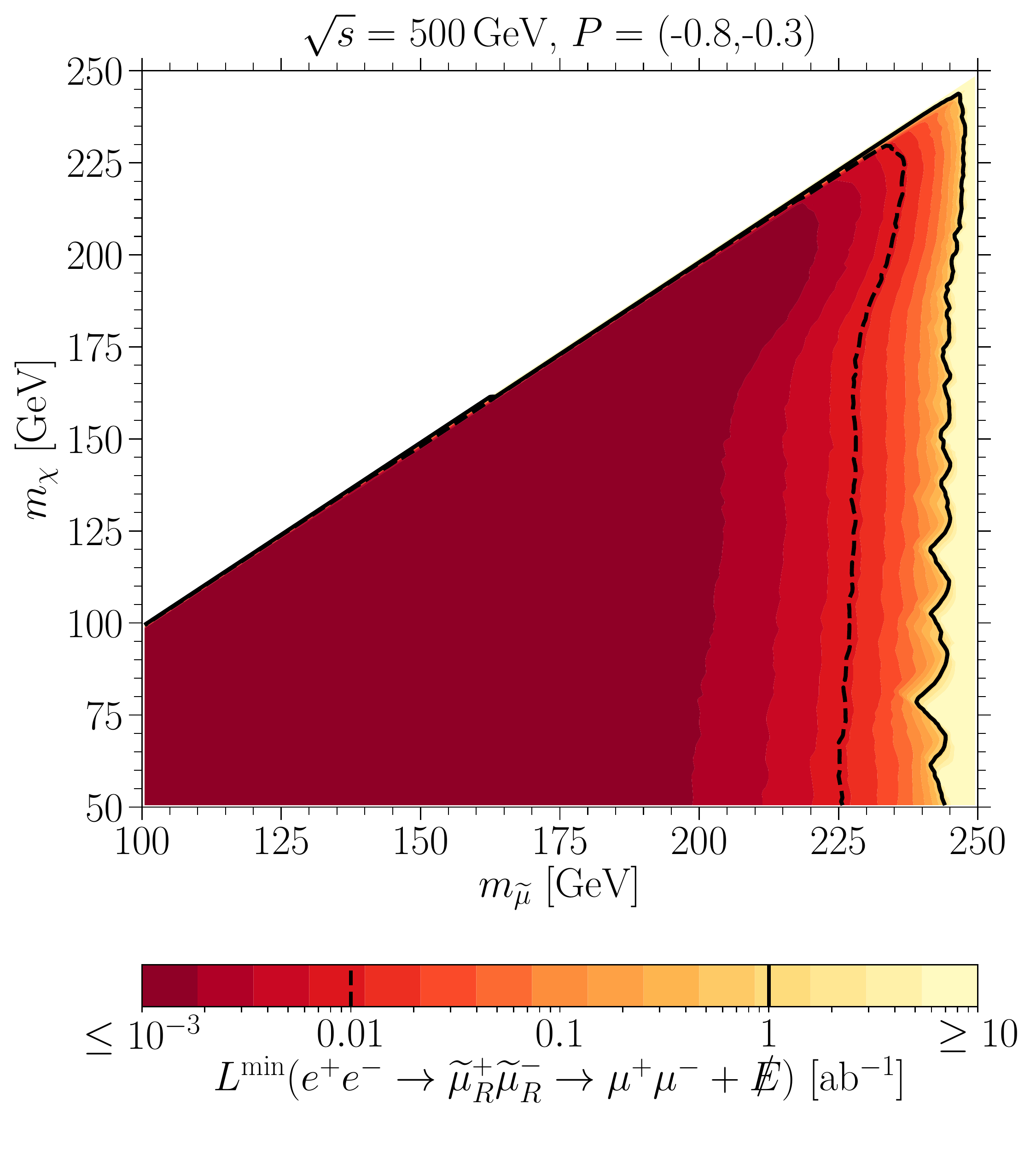}
   \includegraphics[width=0.49\linewidth]{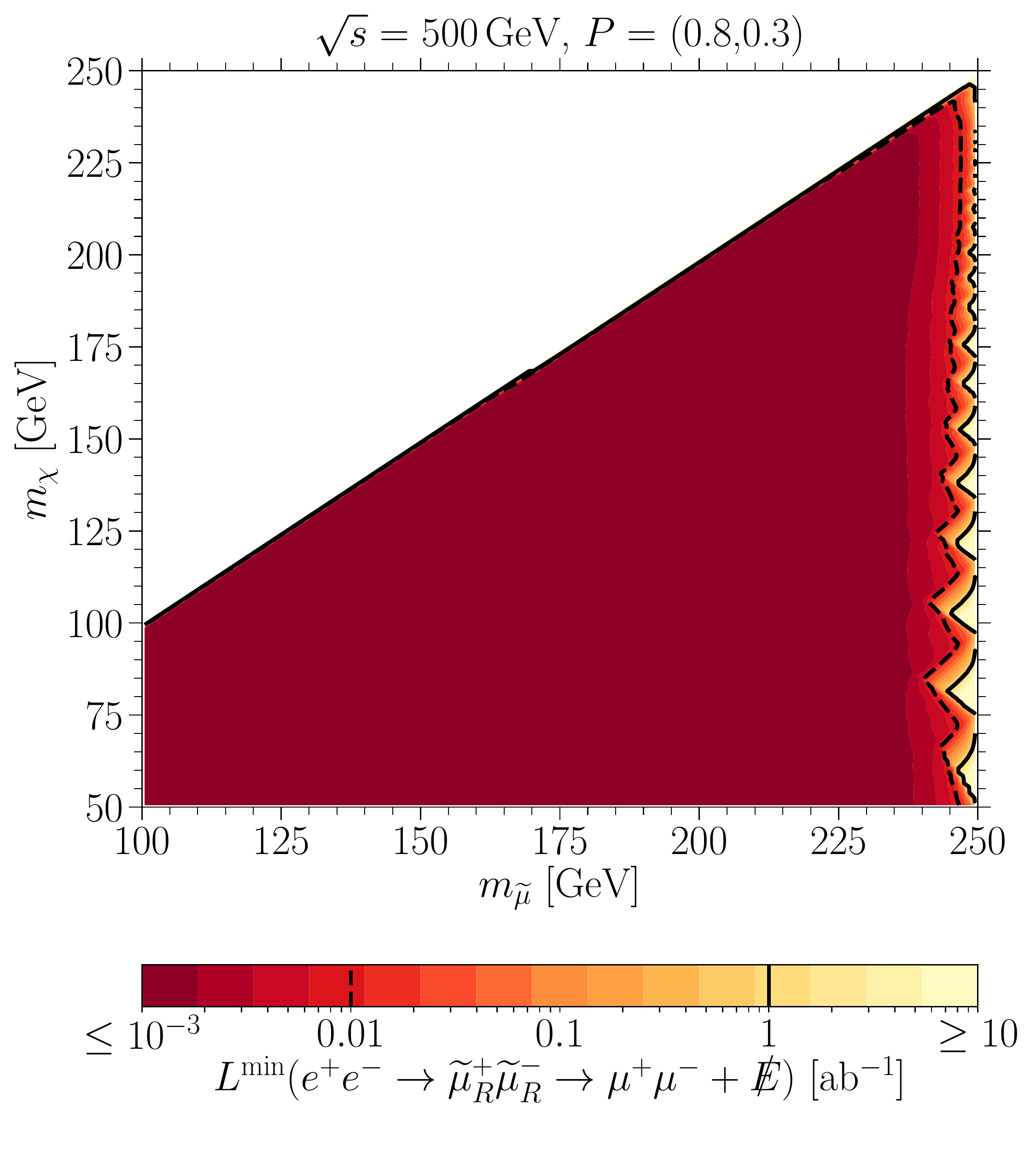}
   \caption{Smallest integrated luminosity $L^{\rm min}$ which would allow to probe the pair production of right-handed smuons at an $e^+e^-$ collider with $\sqrt{s} = 500\,$GeV. The left panel is for mostly left-handed polarization of the electron and positron beams, $P=(-0.8,-0.3)$, while the right panel is for mostly right-handed beam polarization $P=(0.8,0.3)$. Here, we assumed that the smuon can only decay into a muon and a neutralino, ${\rm BR}(\smu^\pm \to \mu^\pm \chi) = 1$, cf. the discussion in Sec.~\ref{sec:model}. Note that the value of the smuon pair production cross section is fixed by the gauge couplings of the smuons, hence, the production cross section in our model discussed in Sec.~\ref{sec:model} is the same as the cross section in SUSY models if the smuon and the neutralino are the next-to-lightest and lightest supersymmetric particles, respectively. The dashed (solid) black line indicates $L^{\rm min} = 10\,{\rm fb}^{-1}$ ($L^{\rm min} = 1\,{\rm ab}^{-1}$).}
   \label{fig:lumi_smuon_500}
\end{figure*}

In Fig.~\ref{fig:reach_500} we show the sensitivity in terms of the smallest cross section, $\sigma^{\rm min}$, which could be probed for smuons in the ($\mu^+ \mu^- + \ME$) final state (left panel) and staus in the ($e^\pm \mu^\mp + \ME$) final state (right panel) for mostly right-handed polarization of the electron and positron beam, $P = (0.8, 0.3)$. 

\begin{figure*}
   \includegraphics[width=0.49\linewidth]{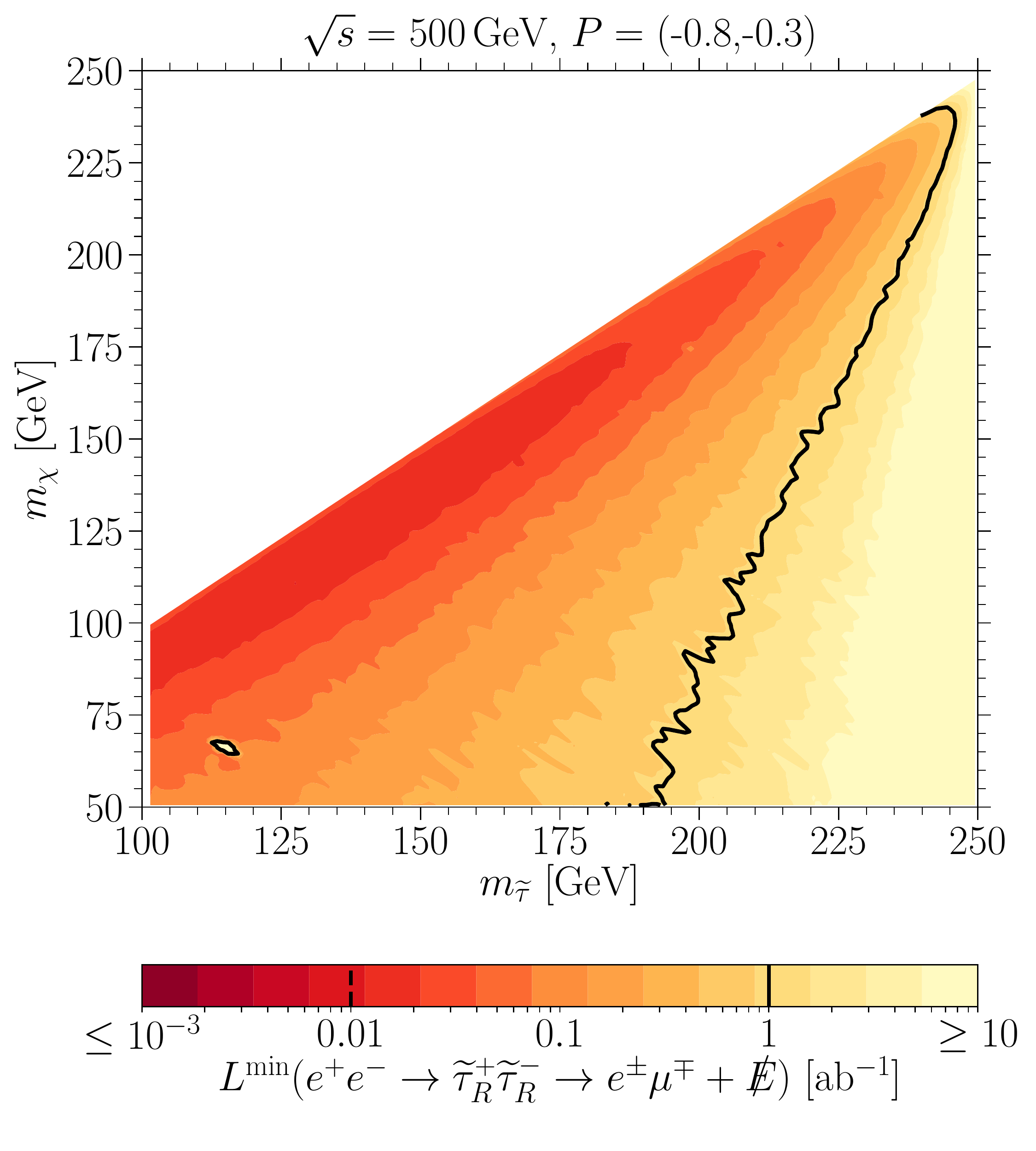}
   \includegraphics[width=0.49\linewidth]{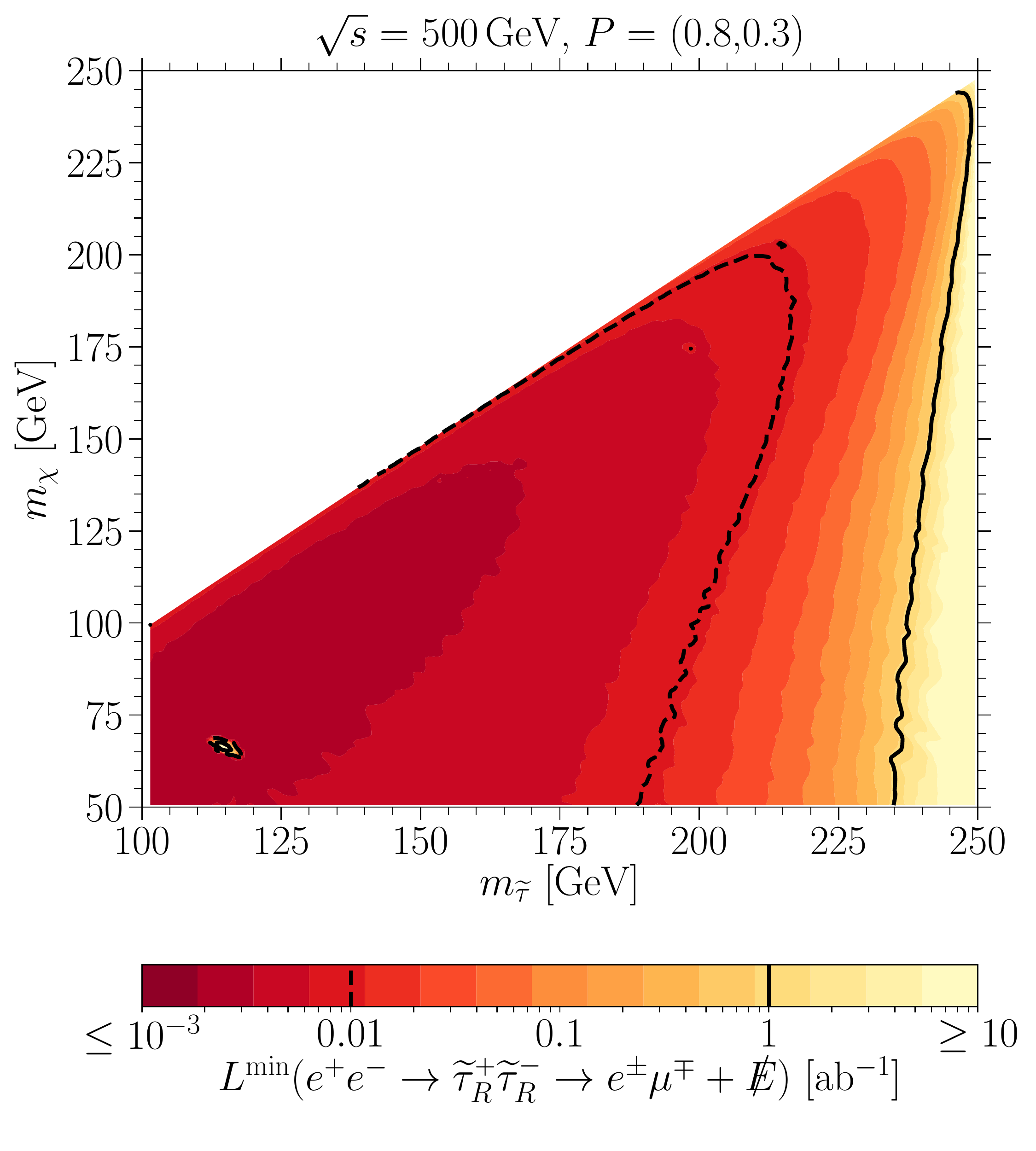}
   \caption{Same as Fig.~\ref{fig:lumi_smuon_500} but for right-handed staus in the ($e^\pm \mu^\mp + \ME$) final state. The ``islands'' at $m_{\stau} \approx 115\,$GeV, $m_\chi \approx 65\,$GeV are an artifact of the statistical fluctuations in the Monte Carlo simulations of the signal (and to a lesser extent, the backgrounds).}
   \label{fig:lumi_stau_emu_500}
\end{figure*}

The sensitivity to smuons in the ($\mu^+ \mu^- + \ME$) final state is best for relatively compressed spectra when the ratio of the neutralino and the smuon mass takes values $m_\chi/m_{\smu} \sim 0.9$. For such mass spectra, the distribution of the kinematic variables we use to separate signal from background, in particular the $p(\ell_1)$ distribution after kinematic cuts, is quite different from the background and cross sections as small as $\sigma^{\rm min} = \mathcal{O}(0.1)\,$fb could be probed at a $\sqrt{s} = 500\,$GeV collider with $L = 1\,{\rm ab}^{-1}$ of data. For more compressed spectra, i.e. $m_\chi/m_{\smu} \gtrsim 0.9$, the charged leptons become too soft, making them more difficult to differentiate from the background, in particular ($\gamma\gamma \to \mu^+\mu^-$), cf. Fig.~\ref{fig:smuons}. For less compressed spectra on the other hand, i.e. $m_\chi/m_{\smu} \lesssim 0.9$, the charged leptons become harder, making the kinematic distributions of the signal more similar to those of the backgrounds from gauge boson pair production, in particular from $W^+W^-$. Note that the sensitivity also improves as the smuon mass approaches the kinematic edge for smuon pair production, $m_{\smu} \to \sqrt{s}/2$. In this limit, the smuons are produced at rest and thus the muon momenta approach a monoenergetic distribution
\begin{equation}
   p(\mu^\pm) \xrightarrow[m_{\smu} \to \sqrt{s}/2]{} \frac{\sqrt{s}}{4} \left( 1 - \frac{4 m_\chi^2}{s} \right)^2 \:,
\end{equation}
which is rather easy to distinguish from the smooth $p(\ell_1)$ distribution of the backgrounds.

For stau searches in the ($e^\pm \mu^\mp + \ME$) final state the behavior of $\sigma^{\rm min}$ with the stau and neutralino masses is different from the smuon case in the ($\mu^+ \mu^- + \ME$) final state for two reasons: 1) this final state does not suffer from large backgrounds for very soft leptons because there is no background process analogous to ($\gamma\gamma \to \mu^+ \mu^-$), and 2) the charged leptons from the stau signal are not promptly produced in the ($\smu^\pm \to \mu^\pm \chi$) decay but via an additional decay step, $\stau^\pm \to (\tau^\pm \to e^\pm/\mu^\pm + 2\nu)+\chi$. The effect of the first difference is that this channel is sensitive to very small cross sections $\sigma^{\rm min} \lesssim \mathcal{O}(0.1)\,$fb even for extremely compressed spectra where $m_\chi \to m_{\stau}$. The sensitivity only starts deteriorating as the mass gap becomes too small to produce on-shell taus, $m_{\stau}-m_{\chi} \lesssim m_\tau = 1.8\,$GeV. On the other hand, due to the latter difference 2), the sensitivity to staus does not improve as the stau mass approaches the kinematic edge, $m_{\stau} \to \sqrt{s}/2$. Unlike in the smuon case, the charged leptons measured in the detector are now produced from the tau decays and hence do not become monoenergetic in this limit.

In Fig.~\ref{fig:reach_500} we also indicate where the sensitivity $\sigma^{\rm min}$ is equal to the production cross sections of the respective final states in the simplified slepton model discussed in Sec.~\ref{sec:model} by the solid black and dashed red lines. We stress that the cross section in our model coincides with the corresponding cross section in SUSY models if the slepton and the neutralino are the next-to-lightest and lightest supersymmetric particle, respectively. In the region to the left and below these lines, the production cross section in our model is larger than the projected sensitivity and hence the corresponding combination of slepton and neutralino masses could be probed at a $\sqrt{s} = 500\,$GeV $e^+e^-$ collider. As we can see from Fig.~\ref{fig:reach_500}, with $L = 1\,{\rm ab}^{-1}$ of data, virtually the entire smuon parameter space where $m_{\smu} < \sqrt{s}/2$ and $m_{\chi} < m_{\smu}$ could be probed. Likewise, for staus we find that most of the corresponding parameter space could be probed in the ($e^\pm \mu^\pm + \ME$) final state, although the coverage is not as good as in the smuon case for small values of $m_{\chi}/m_{\stau}$ and $m_{\stau} \to \sqrt{s}/2$. This is due to two reasons: the signal cross section for staus is suppressed with respect to the smuon case because of the leptonic tau branching ratios, $2\;{\rm BR}(\tau^\pm \to \mu^\pm \nu_\tau \nu_\mu) \times {\rm BR}(\tau^\pm \to e^\pm \nu_\tau \nu_e) \approx 1/16$, and for the reasons discussed above, the sensitivity is not enhanced for $m_{\stau} \to \sqrt{s}/2$.

\begin{figure*}
   \includegraphics[width=0.49\linewidth]{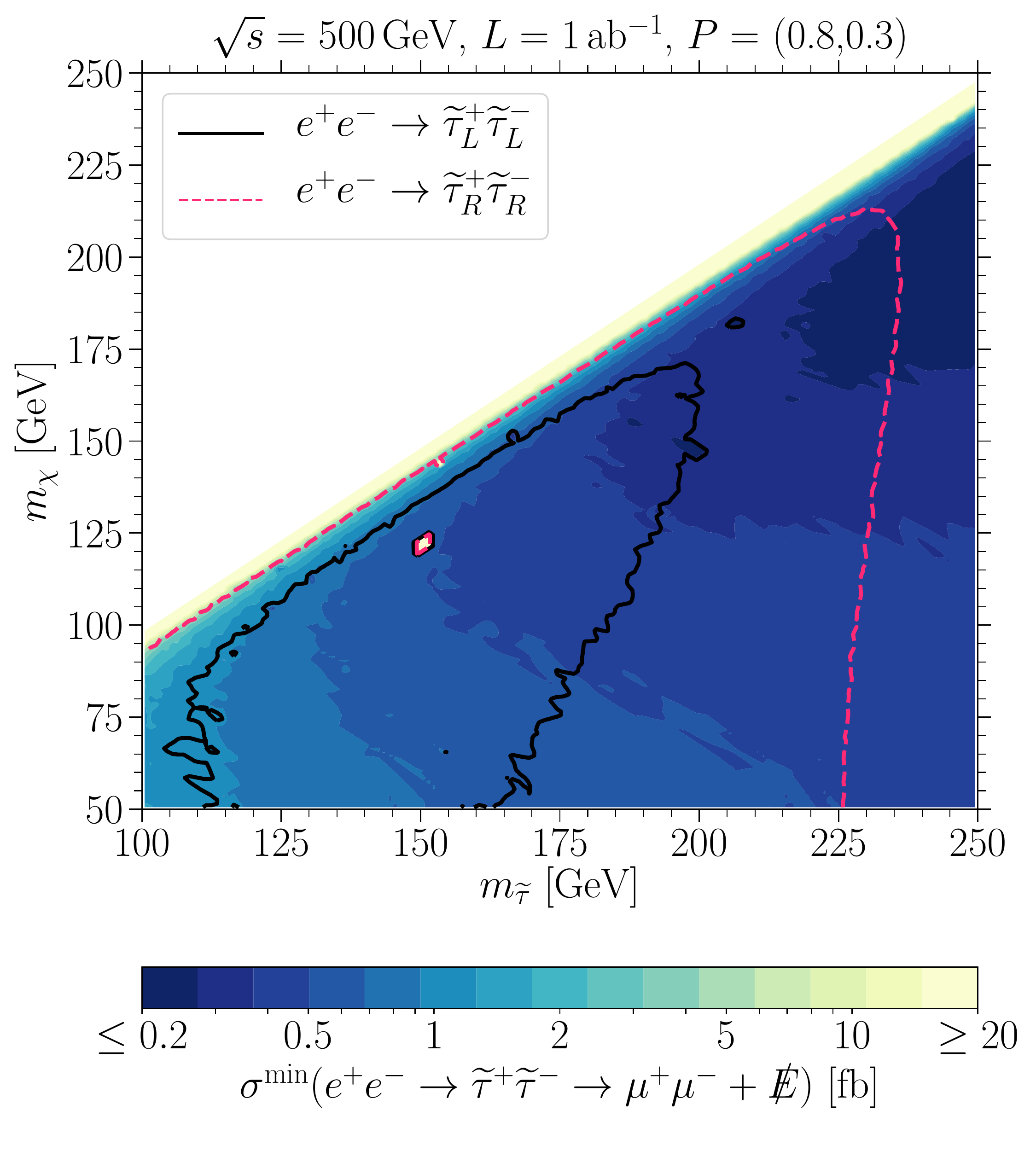}
   \includegraphics[width=0.49\linewidth]{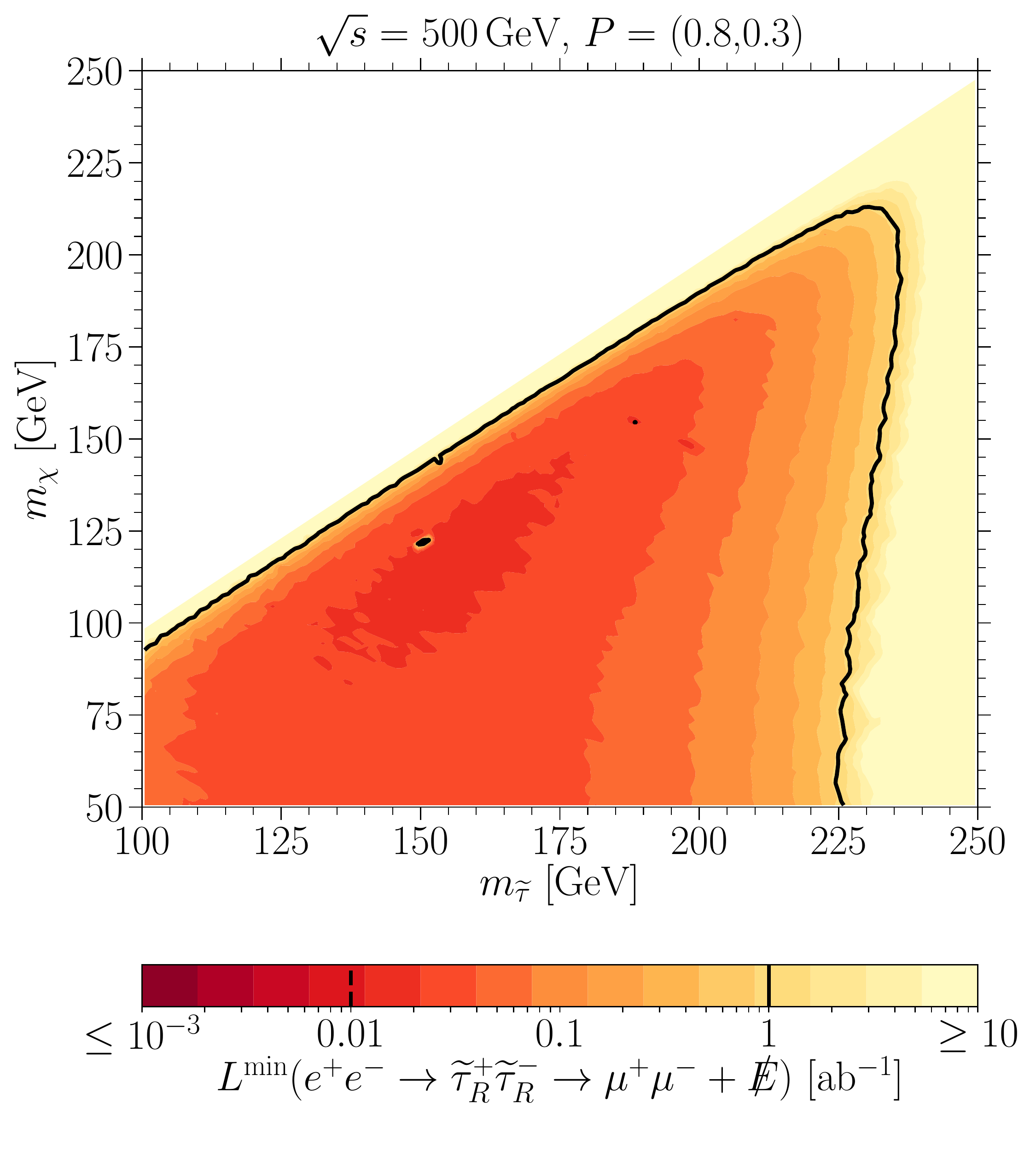}
   \caption{Projected sensitivity to stau pair production in the ($\mu^+\mu^- + \ME$) final state at an $\sqrt{s} = 500\,$GeV $e^+e^-$ collider assuming longitudinal polarization of the electron/positron beams $P = (0.8,0.3)$. {\it Left:} Smallest cross section $\sigma^{\rm min}$ which could be probed with $L = 1\,{\rm ab}^{-1}$ of data. {\it Right:} Smallest integrated luminosity $L^{\rm min}$ required to probe right-handed staus if the signal cross section is given by those in our model described in Sec.~\ref{sec:model}. The ``islands'' at $m_{\stau} \approx 150\,$GeV, $m_\chi \approx 120\,$GeV are an artefact of the statistical fluctuations in the Monte Carlo simulations of the signal (and to a lesser extent, the backgrounds).} 
   \label{fig:reach_lumi_stau_2mu_500}

   \includegraphics[width=0.49\linewidth]{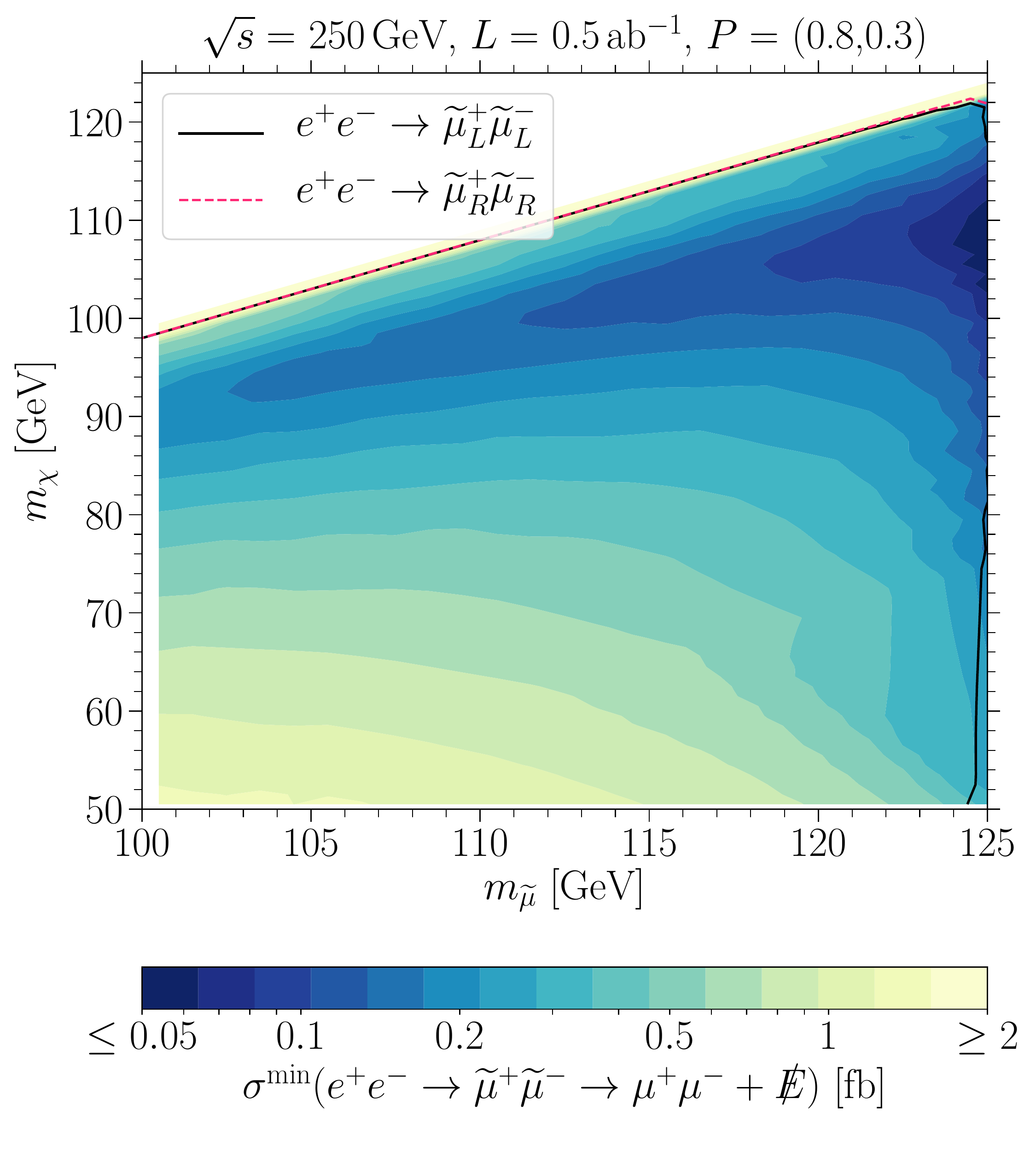}
   \includegraphics[width=0.49\linewidth]{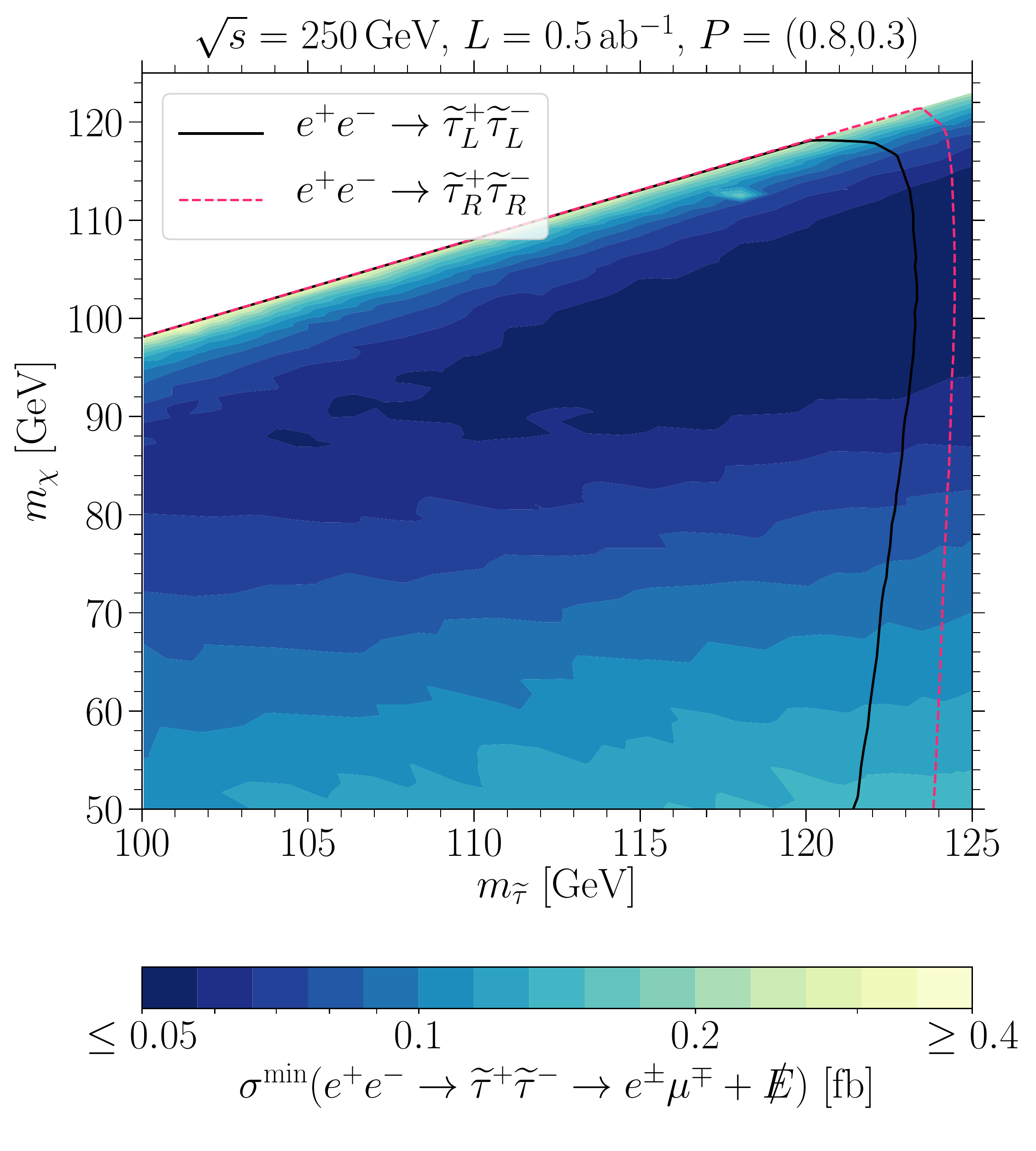}
   \caption{Same as Fig.~\ref{fig:reach_500} but for $\sqrt{s} = 250\,$GeV and $L = 0.5\,{\rm ab}^{-1}$.}
   \label{fig:reach_250}
\end{figure*}

For much of the parameter space, the production cross section of the respective final states in our model is orders of magnitude larger than the smallest cross section we project to be within reach in Fig.~\ref{fig:reach_500}. For example, as long as we are far from the kinematic edge\footnote{As $m_{\smu}^2 \to s/4$ the cross section rapidly falls due to the $\sigma \propto \left( 1 - 4 m_{\sell}^2/s \right)^{3/2}$ scaling, cf. Eq.~\eqref{eq:xsecii}.} for smuon pair production, $m_{\smu}^2 \ll s/4$, the ($e^+e^- \to \smu^+\smu^- \to \mu^+\mu^-\chi\chi$) production cross section takes values of order $\sigma \sim 100\,$fb for $\sqrt{s} = 500\,$GeV. It is thus interesting to ask how much luminosity would be required to probe most of the slepton parameter space at a future $e^+e^-$ collider. 

In Fig.~\ref{fig:lumi_smuon_500} we show the smallest integrated luminosity $L^{\rm min}$ which would probe the pair production of right-handed smuons at an $e^+e^-$ collider with $\sqrt{s} = 500\,$GeV for two different configurations of the beam polarizations, $P=(-0.8,-0.3)$ and $P=(0.8,0.3)$, where the former is mostly left-handed and the latter is mostly right-handed. We cut the color scale off at integrated luminosities $L^{\rm min} = 10^{-3}\,{\rm ab}^{-1}$. Future $e^+e^-$ colliders are envisaged to operate with instantaneous luminosities $\mathscr{L} \gtrsim \mathcal{O}(100)\,{\rm fb}^{-1}\,{\rm yr}^{-1}$~\cite{Fujii:2017vwa,Aihara:2019gcq,Abada:2019zxq,CEPCStudyGroup:2018ghi,Charles:2018vfv}. Thus, proposed $e^+e^-$ colliders would collect $L = 10^{-3}\,{\rm ab}^{-1} = 1\,{\rm fb}^{-1}$ of data within a few days of operations. Looking at Fig.~\ref{fig:lumi_smuon_500}, we see that a few days of running a future $e^+e^-$ collider at $\sqrt{s} = 500\,$GeV would suffice to detect or exclude smuons for most of the accessible parameter space $m_{\smu} < \sqrt{s}/2$, even for very compressed spectra where $m_\chi \to m_{\smu}$.

Comparing the two panels of Fig.~\ref{fig:lumi_smuon_500} exemplifies the effect of the longitudinal polarization of the electron and positron beams on the sensitivity. The mostly right-handed polarization configuration of the beams $P = (0.8,0.3)$ allows coverage of a much larger portion of the parameter space for right-handed smuons than the mostly left-handed polarization configuration $P = (-0.8,-0.3)$. While some sensitivity can be recovered by increasing the integrated luminosity, we see that using an advantageous polarization configuration for a given signal can reduce the luminosity required to cover a particular benchmark point by orders of magnitude. Translated into running time, this means that for the case of right-handed smuon pair production, running for a few days with $P = (0.8,0.3)$ yields better coverage of the parameter space than running for a few years with $P = (-0.8,-0.3)$.

The behavior displayed by the two different panels in Fig.~\ref{fig:lumi_smuon_500} is a rather extreme example of the effects of beam polarization. As discussed in Sec.~\ref{sec:BP_smuon}, the sensitivity to left-handed smuons is much less dependent on the beam polarization because in that case the signal cross section and those of the most relevant backgrounds scale similarly with the beam polarization. The sensitivity to left-handed smuons or smuons with non-vanishing left-right mixing angle would lie in between the two cases shown in Fig.~\ref{fig:lumi_smuon_500}.

In Fig.~\ref{fig:lumi_stau_emu_500} we show the smallest integrated luminosity $L^{\rm min}$ required to probe right-handed staus at a $\sqrt{s} = 500\,$GeV $e^+e^-$ collider in the ($e^\pm \mu^\mp + \ME$) final state. As for smuons in Fig.~\ref{fig:lumi_smuon_500}, we assume that the signal cross section is given by the model described in Sec.~\ref{sec:model}. Recall that the stau pair production cross section is fixed by gauge couplings, and in our model the only allowed decay channel is ${\rm BR}(\stau^\pm \to \tau^\pm \chi) = 1$. For given values of the slepton mass $m_{\sell}$, the signal cross section into the ($e^\pm \mu^\mp + \ME$) final state for staus is thus the same as for smuons into ($\mu^+ \mu^- + \ME$), except for the suppression by the leptonic branching ratios of the taus, $2 \; {\rm BR}(\tau^\pm \to \mu^\pm \nu_\mu \nu_\tau) \times {\rm BR}(\tau^\pm \to e^\pm \nu_e \nu_\tau) \approx 1/16$. While the backgrounds in the ($e^\pm \mu^\mp + \ME$) final state are also smaller than those in the ($\mu^+ \mu^- + \ME$) final state, cf. Sec.~\ref{sec:BP_stau}, the reduction of the signal cross section as well as the change in kinematic distributions still makes staus somewhat more challenging to probe than smuons. Nonetheless, Fig.~\ref{fig:lumi_stau_emu_500} demonstrates that much of the kinematically accessible stau parameter space could be covered at a future $e^+e^-$ collider with relatively low luminosity. As discussed above for the smuon case, the two panels of Fig.~\ref{fig:lumi_stau_emu_500} bracket rather extreme cases, different choices for the beam polarizations or for the left-right mixing angle (including purely left-handed staus) lie in between these cases.

Finally, in Fig.~\ref{fig:reach_lumi_stau_2mu_500}, we show projections for the smallest cross section $\sigma^{\rm min}$ for fixed $L=1\,{\rm ab}^{-1}$ and the minimal luminosity $L^{\rm min}$ when fixing the cross section to that in our model for probing staus in the ($\mu^+ \mu^- + \ME$) final state. We use similar, although more aggressive, kinematic cuts as for smuons in the ($\mu^+\mu^- + \ME$) final state (see Tab.~\ref{tab:cuts_stau_2mu} and Fig.~\ref{fig:stau_2mu}). As for smuons in the ($\mu^+\mu^- + \ME$) final state and staus in ($e^\pm\mu^\mp + \ME$) discussed above, we optimize a signal window in the leading lepton momentum $p(\ell_1)$ in addition to these kinematic cuts to determine the sensitivity. Compared to the ($e^\pm \mu^\mp + \ME$) final state for which we showed results in the right panel of Fig.~\ref{fig:reach_500} and in Fig.~\ref{fig:lumi_stau_emu_500}, we find that the ($\mu^+\mu^- + \ME$) final state is less sensitive to staus regardless of the stau and neutralino masses. As discussed in Sec.~\ref{sec:BP_sim}, while the signal distributions are practically unchanged between the two final states (before kinematic cuts), the ($e^\pm \mu^\mp + \ME$) final state suffers from less SM background.

\subsection{$\sqrt{s}= 250\,{\rm GeV}$}

In Fig.~\ref{fig:reach_250}, we show sensitivity projections to smuons in the ($\mu^+ \mu^- + \ME$) final state and staus in the ($e^\pm \mu^\mp + \ME$) final state for a $\sqrt{s} = 250\,$GeV $e^+e^-$ collider with $L = 0.5\,{\rm ab}^{-1}$ of data. We use the same set of kinematic cuts as for the corresponding $\sqrt{s} = 500\,$GeV projections, although with different numerical values for the cuts, as discussed in Appendices~\ref{app:smuon} and~\ref{app:stau}.

The scaling of the smallest signal cross section which could be probed, $\sigma^{\rm min}$, with the masses of the slepton and neutralino is analogous to what we discussed above for the $\sqrt{s} = 500\,$GeV case. For smuons, the sensitivity is best (in terms of $\sigma^{\rm min}$) for $m_\chi/m_{\smu} \sim 0.9$, and close to the kinematic edge for smuon pair production, $m_{\smu} \to \sqrt{s}/2$. For staus, the sensitivity again does not deteriorate as much in the very compressed regime $m_\chi \to m_{\stau}$ as for the smuon case, but $\sigma^{\rm min}$ also does not improve when approaching the kinematic edge $m_{\stau} \to \sqrt{s}/2$. 

Compared to the signal production cross sections of the respective final states in our simplified model from Sec.~\ref{sec:model}, indicated by the solid black and dashed red lines in Fig.~\ref{fig:reach_250}, we again find that almost all of the kinematically accessible parameter space with $m_{\sell} < \sqrt{s}/2$ and $m_\chi < m_{\sell} + m_{\ell}$ would be probed by a future $e^+e^-$ collider. For most of the parameter space, the cross sections in our simplified model are again much larger than the values of $\sigma^{\rm min}$ shown in Fig.~\ref{fig:reach_250}. Hence, most of the kinematically accessible parameter space would be covered with integrated luminosities $L \ll 1\,{\rm ab}^{-1}$, i.e. with data from a few days of running a future $e^+e^-$ collider. 

While we find excellent sensitivity to smuons and staus even for $\sqrt{s} = 250\,$GeV, the parameter space coverage is hamstrung by the center-of-mass energy, leaving viable models with $m_{\sell}\gtrsim 125$ GeV unreachable. 

\section{Discussion and Conclusions} \label{sec:diss}

Scalar partners of the SM charged leptons are interesting examples of new physics which could be realized close to the electroweak scale while evading detection to date. In this paper, we have explored the extent to which future $e^+e^-$ colliders such as ILC, CLIC, CEPC, or the FCC-ee could probe extensions of the SM by said scalar lepton partners and a new neutral fermion. In analogy to SUSY models, we refer to these states as sleptons ($\sell$) and neutralinos ($\chi$), respectively. While SUSY models are well motivated realizations of models with such new states, our results directly apply to non-SUSY models, as discussed in Sec.~\ref{sec:model}.

In this work, we have focused on scalar partners of the SM's muons and taus, which we refer to as smuons and staus, respectively. They could be pair produced at colliders, and subsequently each slepton would decay into the associated SM lepton and a neutralino. Since the neutralino would leave the detector without depositing energy, the collider signature for smuons is a pair of muons of opposite charge accompanied with missing energy, ($\mu^+\mu^- + \ME$). For staus, a number of final states is possible depending on the decay mode of the taus produced in the stau decays. Here, we focused on leptonic tau decays. Considering SM backgrounds, the most promising final state from stau pair production is ($e^\pm \mu^\mp + \ME$). 

In order for a pair of on-shell sleptons to be produced at an $e^+e^-$ collider, the slepton's mass, $m_{\sell}$, must be smaller than half the center-of-mass energy of the collider, $\sqrt{s}$. Furthermore, for the slepton to be kinematically allowed to decay into a neutralino $\chi$ and a charged SM lepton $\ell$ of the same flavor as the slepton, the respective masses must satisfy $m_{\sell} > m_\chi + m_\ell$. In this work we have shown that future $e^+e^-$ colliders could detect or rule out smuons or staus for most of the kinematically accessible parameter space (where $m_{\sell} < \sqrt{s}/2$) with only a few days of data. This holds regardless of the mass gap $m_{\sell} - m_\chi$. This situation should be contrasted with the sensitivity of hadron colliders. With the data expected to be collected in future runs, the LHC could probe sleptons up to masses of $m_{\sell} \lesssim 500\,$GeV, but only for sizable mass gaps $(m_{\sell} - m_\chi) \gtrsim 60\,$GeV. Intermediate mass gaps are extremely challenging to probe at the LHC, while sensitivity can be regained for relatively (but not too) small mass gaps $m_{\sell} - m_{\chi} \sim 10\,$GeV for slepton masses up to $m_{\sell} \lesssim 200\,$GeV.

In this study, we computed the sensitivity for a future $e^+e^-$ collider operating at $\sqrt{s} = 250\,$GeV and $\sqrt{s} = 500\,$GeV. Our results suggest that a machine with significantly larger center-of-mass energy, e.g. CLIC, should likewise be able to probe sleptons up to $m_{\sell} \lesssim \sqrt{s}/2$ with a few fb$^{-1}$ of data. 

One particular motivation for BSM models containing scalar lepton partners and stable neutral fermions is that models of this type can explain the dark matter in the Universe. For example, dark matter scenarios where either a smuon~\cite{Fukushima:2014yia} or a stau~\cite{Ellis:1999mm,Buckley:2013sca,Pierce:2013rda} is the next-to-lightest and a neutral fermion is the lightest new particle are well-motivated and can lead to distinctive signatures in direct and indirect dark matter searches. A future $e^+e^-$ collider with $\sqrt{s} = 500\,$GeV could detect or rule out most of the feasible parameter space of the incredible bulk, which requires $m_{\sell} \lesssim 150\,$GeV, with only a few fb$^{-1}$ of data. While a dedicated study would be required to fully investigate the coverage of co-annihilation scenarios at higher-energy colliders, our results indicate that an $e^+e^-$ collider with a center of mass energy $\sqrt{s} \simeq 3\,$TeV, such as CLIC, will likely be able to fully probe stau co-annihilation scenarios in the MSSM.

We hope that this paper sheds light on the potential for discovery of new physics with a future $e^+e^-$ collider. Electroweak precision physics, including measuring the properties of the observed Higgs boson to high precision, is a guaranteed and critical physics case for such a collider. Here, we have presented a broad class of models for which a future lepton collider would be capable of directly (and immediately) discovering new physics that would evade detection at the LHC.

\begin{acknowledgments}
We are indebted to Fei~Teng for collaboration in the early stages of this project.
We thank Howard~Baer, James~Brau, Katherine~Freese, Jason~Kumar, and Xerxes~Tata for useful and encouraging discussions.
S.~Baum and P.~Stengel thank the University of Utah, where part of this work was completed, for hospitality.
S.~Baum is supported in part by NSF Grant PHY-1720397, DOE HEP QuantISED award \#100495, and the Gordon and Betty Moore Foundation Grant GBMF7946. 
S.~Baum and P.~Stengel acknowledge support by the Vetenskapsr\r{a}det (Swedish Research Council) through contract No. 638-2013-8993 and the Oskar Klein Centre for Cosmoparticle Physics. 
The work of P.~Sandick is supported by NSF grant PHY-1720282.
\end{acknowledgments}

\appendix
\section{Kinematic cuts: ($\mu^+\mu^- + \ME$)} \label{app:smuon}

In this Appendix we present a more detailed discussion of the kinematic cuts used for event selection in the ($\mu^+\mu^- + \ME$) final state. There are a number of SM processes which give rise to backgrounds in this final state. We simulate exclusive samples of the final states:
\begin{itemize}
   \item ($e^+ e^- \to \mu^+ \mu^-$), $N = 10^7$;
   \item ($e^+ e^- \to \mu^+ \mu^- + 2 \nu$) $N = 10^7$;
   \item ($e^+ e^- \to \mu^+ \mu^- + 4 \nu$), $N = 5 \times 10^6$;
   \item ($\gamma\gamma \to \mu^+ \mu^-$), $N = 10^9$;
   \item ($\gamma\gamma \to \mu^+ \mu^- + 2 \nu$) $N = 10^7$;
   \item ($\gamma\gamma \to \mu^+ \mu^- + 4 \nu$), $N = 10^7$;
\end{itemize}
where $\nu$ denotes a neutrino or antineutrino of any flavor. The number of events $N$ we simulate for the respective final state is chosen to yield sufficiently well-sampled kinematic distributions of the backgrounds after kinematic cuts. In particular, the number of events we simulate for ($\gamma\gamma \to \mu^+ \mu^-$) processes is much larger than for the other background categories because, as we will see, ($\gamma\gamma \to \mu^+\mu^-$) events surviving the kinematic cuts stem from the kinematic tails of the distribution. Hence, we need to oversample this background before kinematic cuts.

Because of the similarities of the kinematic distributions to signal events, we show ($e^+e^- \to W^+ W^- \to \mu^+\mu^- + 2\nu$) events with an on-shell pair of $W^+W^-$ bosons separately from the other ($e^+ e^- \to \mu^+\mu^- + 2\nu$) backgrounds when plotting background distributions. Such events are identified post simulation from the ($e^+e^- \to \mu^+\mu^- + 2\nu$) background samples using the simulation output at the truth level, i.e. before the detector simulation.

In addition to the background processes, we simulate signal samples for a grid of combinations of the slepton and neutralino masses, both for smuon pair production, ($e^+e^- \to \smu^+ \smu^- \to \mu^+\mu^-\chi\chi$) and for stau pair production, [$e^+e^- \to \stau^+\stau^- \to (\tau^+ \to \mu^+ \nu_\mu \bar{\nu}_\tau) + (\tau^- \to \mu^- \bar{\nu}_\mu \nu_\tau) + \chi\chi$]. When showing kinematic distributions of the signal compared to the background, we select a set of benchmark values for the slepton and neutralino mass, while the sensitivity projections presented in Sec.~\ref{sec:sens} make use of the full grid of mass combinations. 

We use a set of kinematic observables for our event selection:
\begin{itemize}
   \item the momentum of the leading (charged) lepton, $p(\ell_1)$,
   \item the (cosine of the) polar angle of the leading lepton with respect to the beam axis, $\cos\theta(\ell_1)$,
   \item the invariant mass of the charged lepton system, $m_{\ell\ell}$,
   \item the opening angle between the charged leptons in the plane transverse to the beam axis, $\Delta\phi(\ell^+,\ell^-)$,
   \item the missing energy, $\ME$,
   \item the (cosine of the) polar angle of $\ME$ with respect to the beam axis, $\cos\theta(\ME)$.
\end{itemize}

\begin{figure*}
   \includegraphics[width=0.49\linewidth]{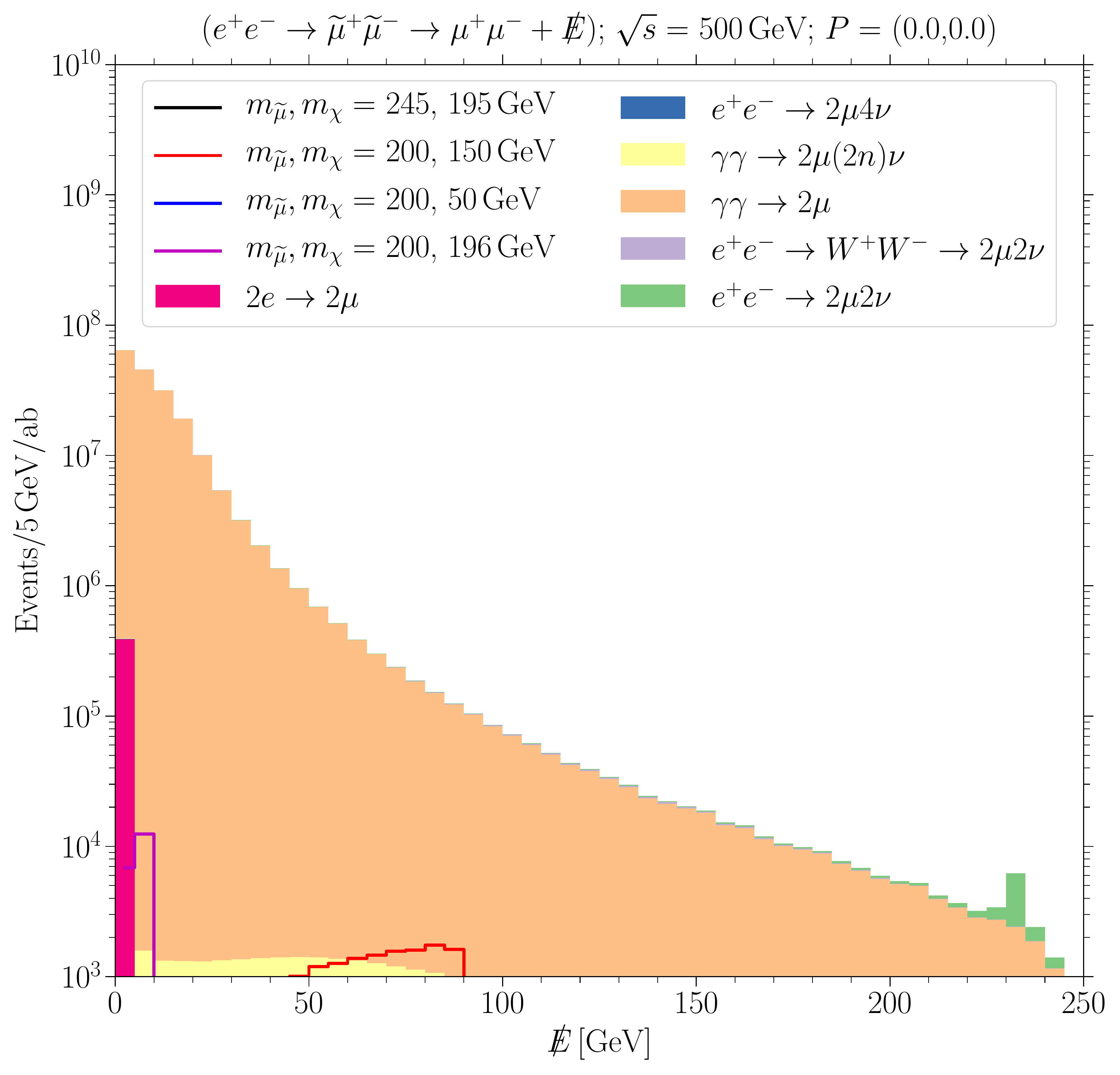}
   \includegraphics[width=0.49\linewidth]{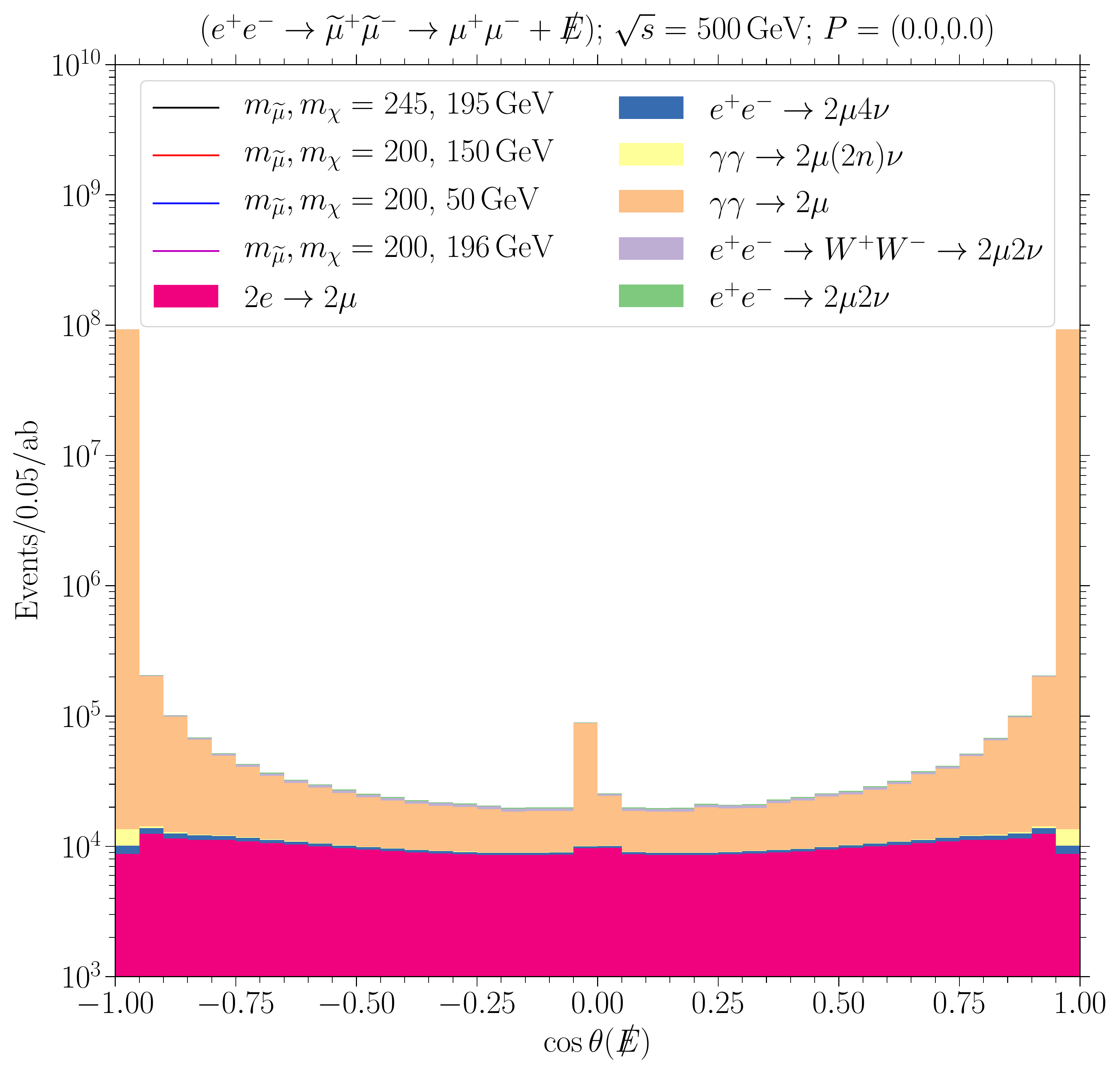}
   \caption{Distribution of the missing energy $\ME$ (left) and its polar angle $\cos\theta(\ME)$ (right) in the $(\mu^+ \mu^- + \ME)$ final state before any kinematic cuts. The stacked histogram shows the different background components as listed in the legend. The solid lines show $(e^+e^- \to \gamma/Z \to \smu^+ \smu^- \to \mu^+\mu^-\chi\chi)$ signal distributions for various benchmark values of the smuon mass $m_{\smu}$ and neutralino mass $m_\chi$ indicated in the legend.}
   \label{fig:smuon_nocuts}

   \includegraphics[width=0.49\linewidth]{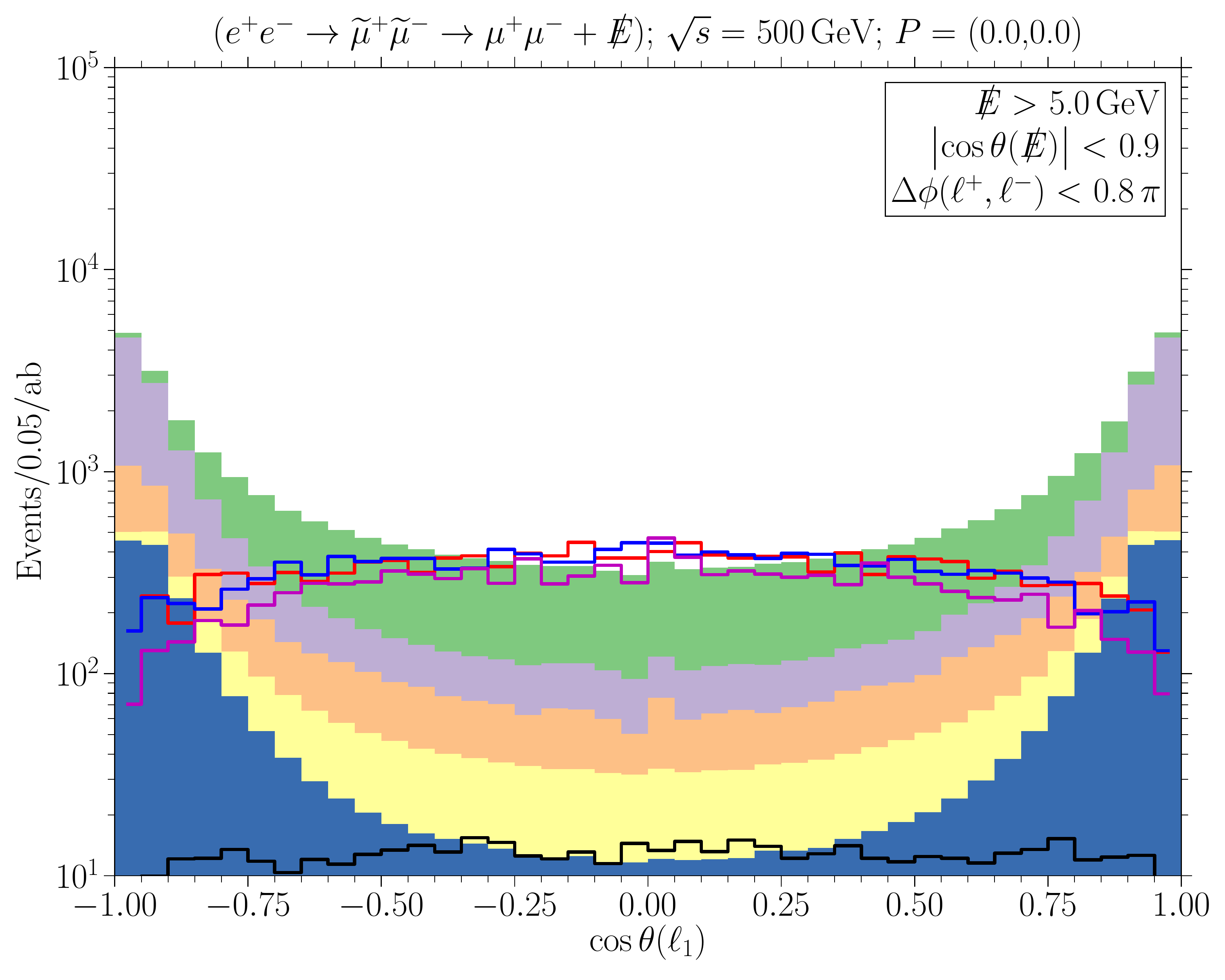}
   \includegraphics[width=0.49\linewidth]{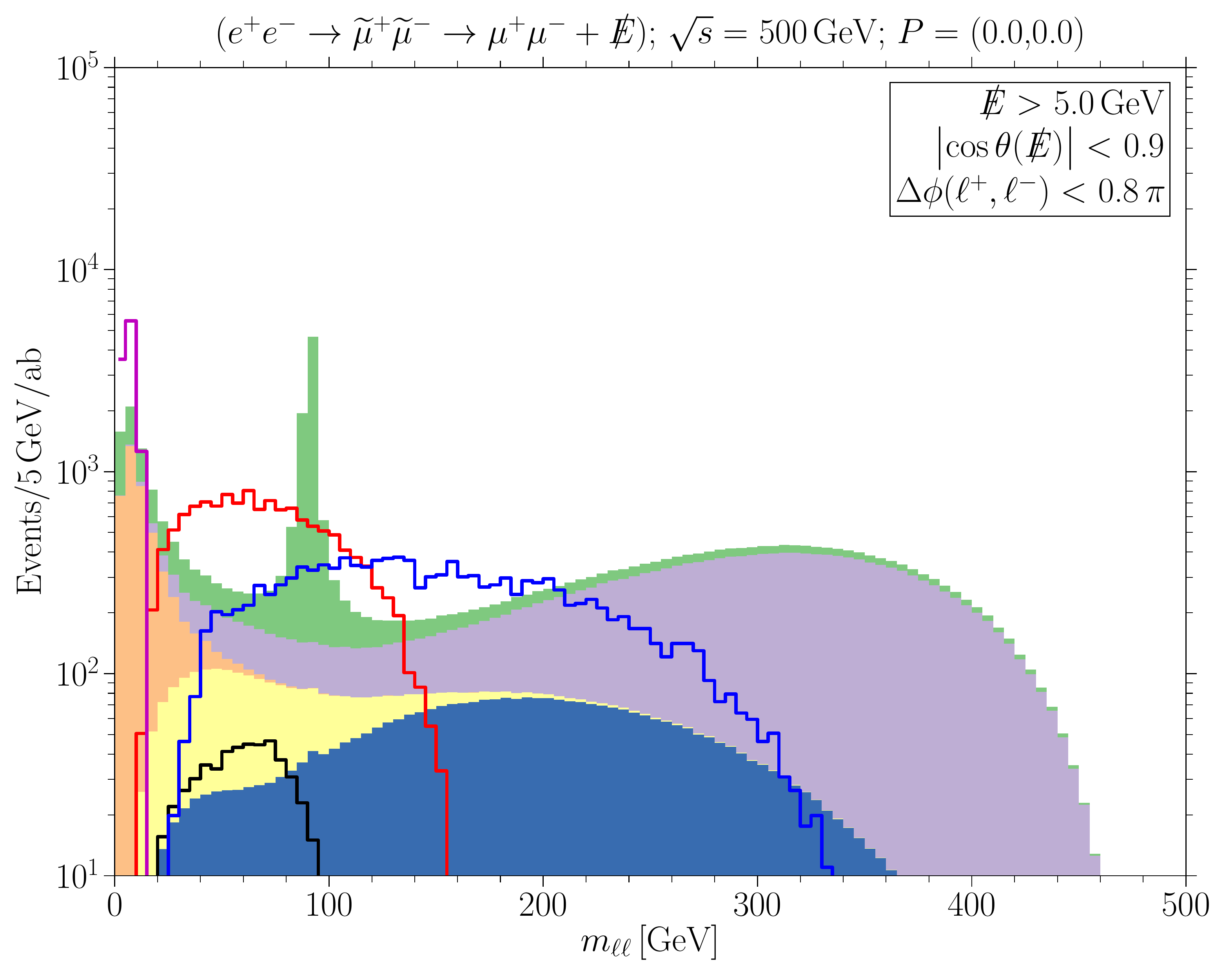}
   \caption{Distributions of the polar angle of the leading charged lepton $\cos\theta(\ell_1)$ (left) and the invariant mass of the charged leptons $m_{\ell\ell}$ (right) in the $(\mu^+ \mu^- + \ME)$ final state after the first set of kinematic cuts described in the text. We use the same colors for the respective background components and signal benchmarks as in Fig.~\ref{fig:smuon_nocuts}.}
   \label{fig:smuon_cut0}
\end{figure*}

In the following, we describe the kinematic cuts used for event selection for smuon searches at $\sqrt{s} = 500\,$GeV. The kinematic cuts we use for $\sqrt{s} = 250\,$GeV differ only in the numeric values chosen for the kinematic cuts, cf. Tab.~\ref{tab:cuts_smu}. We will briefly comment on our selection of kinematic cuts for stau searches in the ($\mu^+\mu^- + \ME$) final state at the end of this section; the numerical values of the corresponding kinematic cuts are listed in Tab.~\ref{tab:cuts_stau_2mu}.

In Fig.~\ref{fig:smuon_nocuts} we show the distribution of $\ME$ and $\cos\theta(\ME)$ for the backgrounds together with those for various benchmark signal distributions before kinematic cuts. The background distributions are dominated by ($e^+e^- \to \mu^+ \mu^-$) and ($\gamma\gamma \to \mu^+\mu^-$) events because of the large corresponding production cross sections. While the signal events as well as background events with neutrinos in the final state do feature true missing transverse energy, this is not the case for ($e^+e^-/\gamma\gamma \to \mu^+\mu^-$) backgrounds. For ($e^+e^- \to \mu^+\mu^-$) events, $\ME > 0$ arises only from mis-measurements of the muon momenta in the detector. ($\gamma\gamma \to \mu^+\mu^-$) events on the other hand do feature true $\ME$, because the center-of-mass frame of the photons in the initial state does not coincide with the detector frame, but is boosted along the beam axis. However, such events do not feature true missing transverse energy. Thus, we require events to satisfy $\ME > 5\,$GeV, $\left|\cos\theta(\ME)\right| < 0.9$, and $\Delta\phi(\ell^+,\ell^-) < 0.8\,\pi$. Effectively, these kinematic cuts amount to requiring events to feature non-vanishing missing transverse energy. 

In Fig.~\ref{fig:smuon_cut0} we show the $\cos\theta(\ell_1)$ and $m_{\ell\ell}$ distributions after this first series of kinematic cuts. Comparing to Fig.~\ref{fig:smuon_nocuts}, we see that the ($e^+e^-/\gamma\gamma \to \mu^+\mu^-$) backgrounds are much reduced, and the largest remaining backgrounds are ($e^+e^- \to \mu^+\mu^- + 2\nu$) [including ($e^+e^- \to W^+W^- \to \mu^+\mu^- + 2\nu$)] events.

We use a second set of kinematic cuts to further reduce the number of background events: From the $\cos\theta(\ell_1)$ distributions we can note that the backgrounds distributions peak at $\left|\cos\theta(\ell_1)\right| \to 1$, while the signal distributions are largest at $\cos\theta(\ell_1) \to 0$. This behavior is due to the different physical processes giving rise to the respective distributions: Many of the background processes proceed via $t$-channel exchange of particles. Most importantly, one of the largest production channels for $W^+W^-$ pairs at $e^+e^-$ colliders is the $t$-channel exchange of a neutrino. Likewise, one of the largest contribution to ($\gamma\gamma \to \mu^+\mu^-$) production stems from $t$-channel exchange of a muon. Signal events on the other hand are produced via $s$-channel exchange of a photon or $Z$-boson, cf. Fig.~\ref{fig:diagram}. The signal-to-background ratio is thus enhanced by selecting events with $\left|\cos\theta(\ell_1)\right| < 0.8$, preferring $s$-channel over $t$-channel processes.

In order to reduce backgrounds stemming from [$e^+e^-/\gamma\gamma \to (Z \to \mu^+\mu^-) + (Z \to \nu \bar{\nu})$] events, we exclude events where the invariant mass of the charged lepton system is approximately given by the mass of the $Z$-boson, $m_Z$. 

Finally, signal events contain a pair of massive neutralinos in the final state. Thus, some of the center-of-mass energy of the collision has to go to the production of these neutralinos, limiting the possible momenta of the charged leptons. Background events on the other hand contain only particles with masses negligible compared to $\sqrt{s}$ in the final state. We select events with $m_{\ell\ell} < 300\,$GeV to further enhance the ratio of signal to background events. 

In Fig.~\ref{fig:smuons} we show the $p(\ell_1)$ distributions after these kinematic cuts. When computing the sensitivity in Sec.~\ref{sec:sens}, we optimize a signal window in $p(\ell_1)$ in addition to the kinematic cuts discussed above. From Fig.~\ref{fig:smuons} we can already see that the benchmark points shown are straightforward to probe at a future collider with a few fb$^{-1}$ of data, except close to the kinematic threshold for slepton pair production, $m_{\sell} \to \sqrt{s}/2$, where the slepton pair production cross section falls quickly because of the kinematic suppression, $\sigma \propto \left( 1-4m_{\sell}^2/s\right)^{3/2}$.

\begin{figure}
   \includegraphics[width=\linewidth]{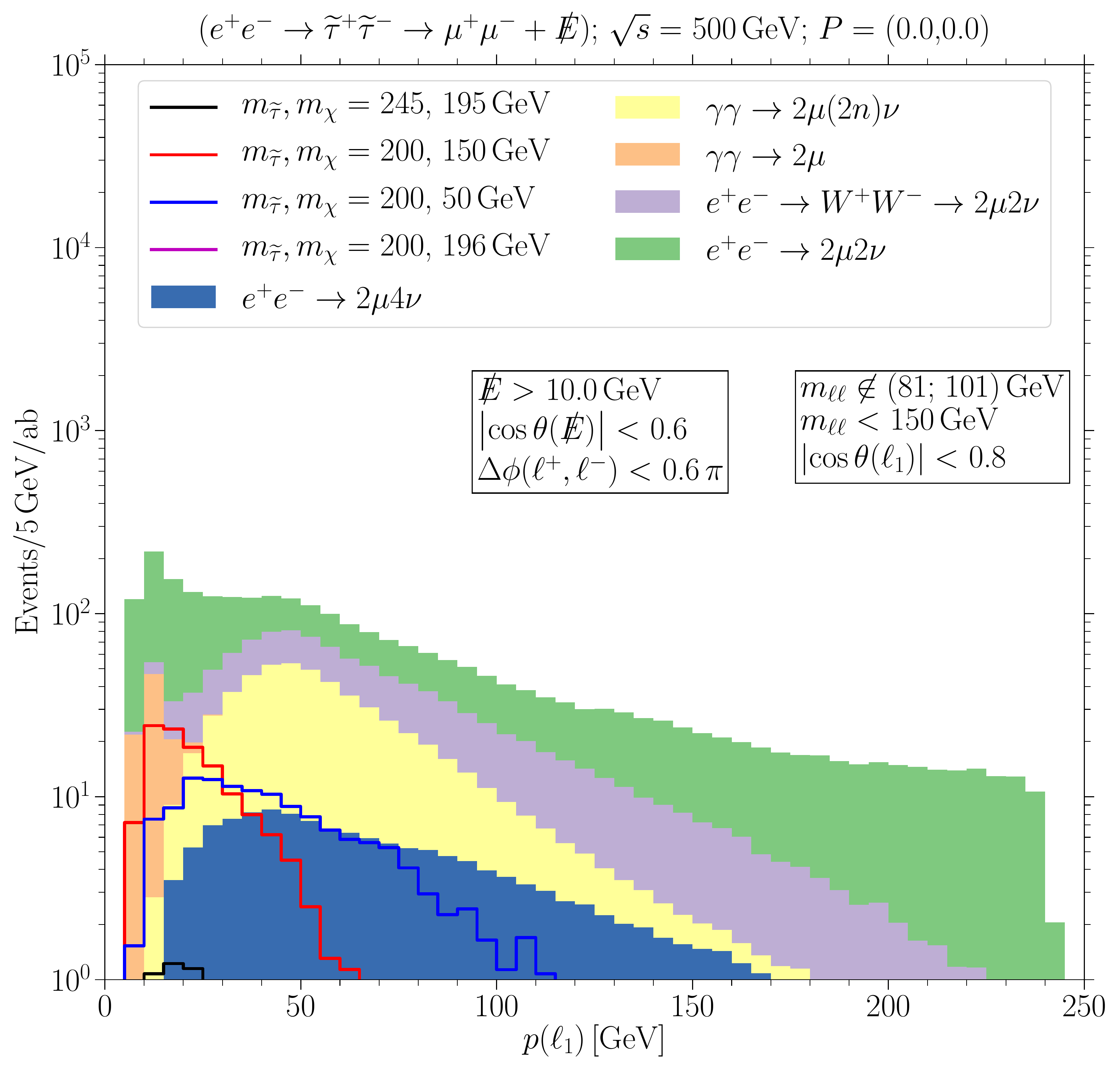}
   \caption{Distribution of leading lepton momentum $p(\ell_1)$ for stau searches in the ($\mu^+\mu^- + \ME$) final state after kinematic cuts.}
   \label{fig:stau_2mu}
\end{figure}

In principle, staus can also be probed in the ($\mu^+\mu^- + \ME$) final state via [$e^+ e^- \to \stau^+ \stau^- \to (\tau^+ \to \mu^+ \nu_\mu \bar{\nu}_\tau) + (\tau^- \to \mu^- \bar{\nu}_\mu \nu_\tau) + \chi\chi$] processes. Compared to the smuon case, probing staus is complicated in two ways: The branching ratio for leptonic tau decays is ${\rm BR}(\tau^\pm \to \mu^\pm \nu_\mu \nu_\tau) \approx 18\,\%$, suppressing the signal cross section, and the additional neutrinos produced in the tau decays lead to softer muons in the final state than those from ($e^+ e^- \to \smu^+\smu^- \to \mu^+\mu^-\chi\chi$). In order to be sensitive to staus in the ($\mu^+\mu^- + \ME$) final state, one can attempt to reduce the background further by using more aggressive kinematic cuts than for smuon searches. Since the charged leptons are softer than for smuons, it is particularly useful to employ a more aggressive upper cut on the invariant mass of the leptons system. 

In Fig.~\ref{fig:stau_2mu} we show the $p(\ell_1)$ distributions for benchmark values of the stau and neutralino masses together with the background distributions after the set of kinematic cuts listed in Tab.~\ref{tab:cuts_stau_2mu}. As we can see, while not impossible, despite the more aggressive kinematic cuts it is much more challenging to search for staus in the ($\mu^+\mu^- + \ME$) final state than for smuons. We show the corresponding sensitivity in the plane of the stau and neutralino masses in Fig.~\ref{fig:reach_lumi_stau_2mu_500}. Fortunately, another final state, ($e^\pm \mu^\mp + \ME$), is available for stau searches at lepton colliders. We discuss the event selection for stau searches in this final state in Appendix~\ref{app:stau}. While the signal cross section as well as kinematic distributions of signal events is virtually the same for staus in the ($e^\pm\mu^\mp + \ME$) and the ($\mu^+\mu^- + \ME$) final states, backgrounds are much lower because fewer SM processes give rise to ($e^\pm \mu^\mp + \ME$) final states than to ($\mu^+\mu^- + \ME$) final states.

\section{Kinematic cuts: ($e^\pm\mu^\mp + \ME$)} \label{app:stau}

\begin{figure*}
   \includegraphics[width=0.49\linewidth]{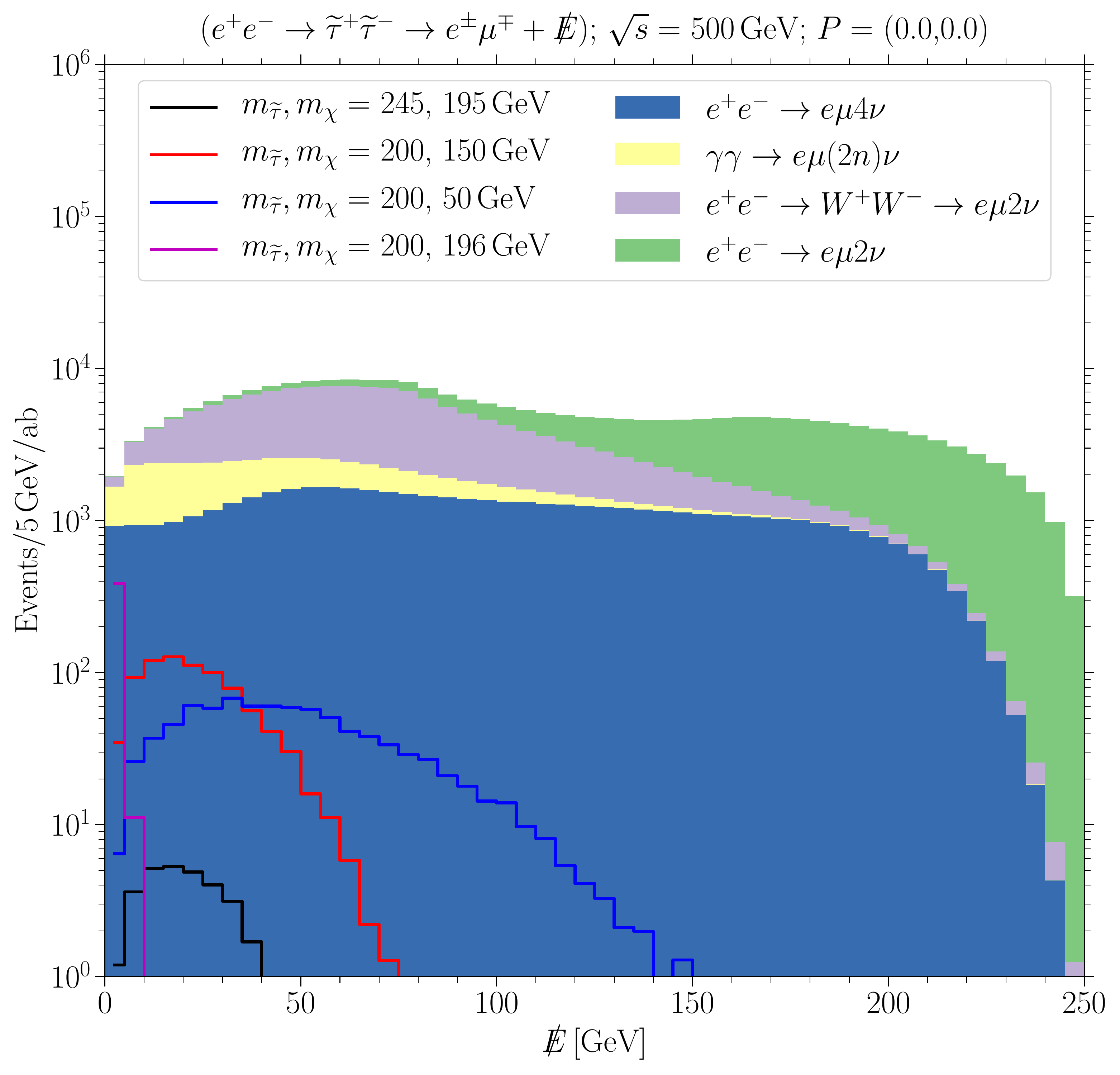}
   \includegraphics[width=0.49\linewidth]{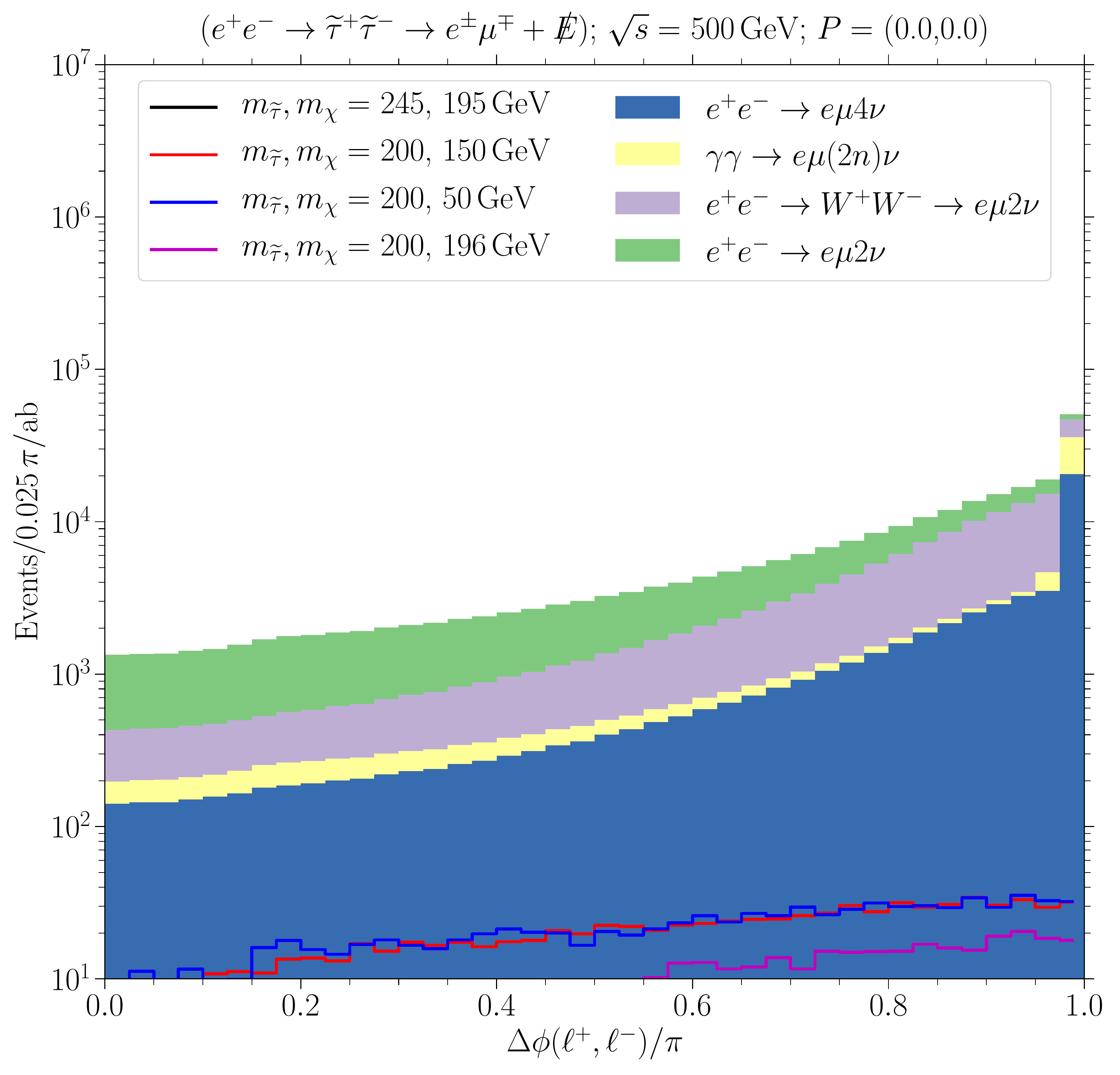}
   \caption{Distribution of the missing energy $\ME$ (left) and the opening angle between the charged leptons in the plane transverse to the beam, $\Delta\phi(\ell^+,\ell^-)$ (right) in the $(e^\pm \mu^\mp + \ME)$ final state before any kinematic cuts. The stacked histogram shows the different background components as listed in the legend. The solid lines show $(e^+e^- \to \gamma/Z \to \stau^+ \stau^- \to e^\pm\mu^\mp\chi\chi)$ signal distributions for various benchmark values of the stau mass $m_{\stau}$ and neutralino mass $m_\chi$ indicated in the legend.}
   \label{fig:stau_emu_nocuts}
\end{figure*}

In this Appendix we present a more detailed discussion of the kinematic cuts used for event selection in the ($e^\pm \mu^\mp + \ME$) final state. In order to account for SM backgrounds, we simulate exclusive samples of the final states:
\begin{itemize}
   \item ($e^+ e^- \to e^\pm \mu^\mp + 2\nu$), $N = 10^7$;
   \item ($e^+ e^- \to e^\pm \mu^\mp + 4\nu$), $N = 5 \times 10^6$;
   \item ($\gamma\gamma \to e^\pm \mu^\mp + 2\nu$), $N = 10^7$;
   \item ($\gamma\gamma \to e^\pm \mu^\mp + 4\nu$), $N = 10^7$;
\end{itemize}
where $\nu$ denotes a neutrino or antineutrino of any flavor, and $N$ denotes the number of events we simulate for the respective final state. Compared to our background simulations in the ($\mu^+\mu^- + \ME$) final state, we do not need to simulate background processes without neutrinos in the final state since flavor conservation forbids ($e^+e^-/\gamma\gamma \to e^\pm \mu^\mp$) production. 

As for the ($\mu^+\mu^- + \ME$) final state, we will show ($e^+e^- \to W^+ W^- \to e^\pm \mu^\mp + 2\nu$) events with an on-shell pair of $W^+W^-$ boson separately from the other ($e^+e^- \to e^\pm \mu^\mp + 2\nu$) backgrounds when plotting signal and background distributions. Such events are identified post-simulation from the ($e^+e^- \to e^\pm \mu^\mp + 2\nu$) background sample using the simulation output at the truth level. 

As for the ($\mu^+\mu^- + \ME$) final state, we simulate signal samples from stau pair production [$e^+e^- \to \stau^+\stau^- \to (\tau^\pm \to e^\pm \nu_e \nu_\tau) + (\tau^\mp \to \mu^\mp \nu_\mu \nu_\tau) + \chi\chi$]) for a grid of combinations of the stau and neutralino masses. When showing kinematic distributions of the signal compared to the background, we select a set of benchmark values for the stau and neutralino mass, while the sensitivity projections presented in Sec.~\ref{sec:sens} make use of the full grid of mass combinations.

We use the same kinematic observables for our event selection as in the ($\mu^+\mu^- + \ME$) final state. Note that we do not differentiate between kinematic variables of the electron and the muon in the final state beyond requiring events to feature exactly one reconstructed electron and one reconstructed muon of opposite charge. 

In the following, we describe the kinematic cuts used for event selection for stau searches at $\sqrt{s} = 500\,$GeV. The kinematic cuts we use for $\sqrt{s} = 250\,$GeV differ only in the numeric values chosen for the kinematic cuts, cf. Tab.~\ref{tab:cuts_stau_emu}. 

In Fig.~\ref{fig:stau_emu_nocuts} we show the $\ME$ and $\Delta\phi(\ell^+,\ell^-)$ distributions for the backgrounds together with those for various benchmark signal distributions before kinematic cuts. Compared to the distributions for the ($\mu^+\mu^- + \ME$) final state, cf. Fig.~\ref{fig:smuon_nocuts}, we can note that due to the absence of the analogue of ($e^+e^-/\gamma\gamma \to \mu^+\mu^-$) backgrounds, there is no need to use a lower cut on $\ME$ or an upper cut on $|\cos\theta(\ME)|$. Because all SM events with exactly one electron and muon of (opposite charge) at $e^+e^-$ colliders feature at least two neutrinos in the final state, all background events have true missing transverse energy.

From the $\ME$ distributions shown in the left panel of Fig.~\ref{fig:stau_emu_nocuts}, we see that the signal distributions are peaked at relatively small values of $\ME$ while the backgrounds remain large for $\ME \to \sqrt{s}/2$. There are two reasons for the smaller $\ME$ of the signal: First, the massive neutralinos in the final state absorb some of the center-of-mass energy, which thus cannot be distributed into the momenta of the states. Second, the signal process contains six states [four (massless) neutrinos and two (massive) neutralinos] in the final state which leave the detector without depositing energy. Since the momenta of these six states will partially cancel, the resulting $\ME$ is smaller than for the backgrounds, where the dominant component features only two neutrinos in the final state. Thus, we use an upper cut $\ME < 50\,$GeV to suppress background events.

In the right panel of Fig.~\ref{fig:stau_emu_nocuts} we show the $\Delta\phi(\ell^+,\ell^-)$ distribution. The backgrounds feature on average larger opening angles between the charged leptons than the signal for the same reasons as why the signal spectra feature rather small $\ME$. When producing only massless neutrinos in addition to the charged leptons, the charged leptons are produced approximately back-to-back if the neutrinos are soft. Thus, we select events satisfying $\Delta\phi(\ell^+,\ell^-) < 0.8\,\pi$. 

Finally, as discussed for smuon searches in the ($\mu^+\mu^- + \ME$) final state, much of the background stems from $t$-channel processes while stau pairs are produced via $s$-channel exchange of a photon or $Z$-boson. Hence, we select events with $\left|\cos\theta(\ell_1)\right| < 0.6$. 

We show the $p(\ell_1)$ distributions after these kinematic cuts in Fig.~\ref{fig:stau}. Similar to the smuon case in the ($\mu^+\mu^- + \ME$) final state, we can immediately deduce that the benchmark points shown are straightforward to probe at future colliders, except close to the kinematic threshold for stau pair production, $m_{\stau} \to \sqrt{s}/2$. Different from the smuon case, most of the constraining power will now come from events with very soft leptons. This is because, as discussed above, stau pair production gives rise to quite soft charged leptons, and SM backgrounds in the ($e^\pm \mu^\mp + \ME$) final state are relatively low at small lepton momenta.

\bibliography{TheBib}

\end{document}